\documentclass[preprint, 3p, sort&compress]{elsarticle}

\usepackage{lineno,hyperref}

\journal{Mechanical Systems and Signal Processing}

\usepackage{amsmath,amsfonts,amssymb}
\usepackage{bm}

\DeclareMathOperator*{\argmin}{arg\,min}
\DeclareMathOperator\erf{erf}
\usepackage[official]{eurosym}
\usepackage[skip=0pt]{caption}
\usepackage[inline]{enumitem}
\usepackage{subcaption}
\usepackage{multicol}
\usepackage{xcolor}
\usepackage{comment}
\begin{document}
	
	\begin{frontmatter}

		\title{A framework for quantifying the value of vibration-based structural health monitoring}
		
		\author[mymainaddress,mytertiaryaddress]{Antonios Kamariotis\corref{mycorrespondingauthor}}
		\cortext[mycorrespondingauthor]{Corresponding author}
		\ead{antonis.kamariotis@tum.de}
		\author[mysecondaryaddress,mytertiaryaddress]{Eleni Chatzi}
		\ead{chatzi@ibk.baug.ethz.ch}
		\author[mymainaddress]{Daniel Straub}
		\ead{straub@tum.de}
		
		\address[mymainaddress]{Engineering Risk Analysis Group, Technical University of Munich, Theresienstrasse 90, 80333 Munich, Germany}
		\address[mysecondaryaddress]{Institute of Structural Engineering, ETH Zurich, Stefano-Franscini-Platz 5, 8093 Zurich, Switzerland}
		\address[mytertiaryaddress]{Institute for Advanced Study, Technical University of Munich, Lichtenbergstrasse 2a, 85748 Garching, Germany}
		
		\begin{abstract}
        The difficulty in quantifying the benefit of Structural Health Monitoring (SHM) for decision support is one of the bottlenecks to an extensive adoption of SHM on real-world structures. In this paper, we present a framework for such a quantification of the value of vibration-based SHM, which can be flexibly applied to different use cases. These cover SHM-based decisions at different time scales, from near-real time diagnostics to the prognosis of slowly evolving deterioration processes over the lifetime of a structure. The framework includes an advanced model of the SHM system. It employs a Bayesian filter for the tasks of sequential joint deterioration state-parameter estimation and structural reliability updating, using continuously identified modal and intermittent visual inspection data. It also includes a realistic model of the inspection and maintenance decisions throughout the structural life-cycle. On this basis, the Value of SHM is quantified by the difference in expected total life-cycle costs with and without the SHM. We investigate the framework through application on a numerical model of a two-span bridge system, subjected to gradual and shock deterioration, as well as to changing environmental conditions, over its lifetime. The results show that this framework can be used as an a-priori decision support tool to inform the decision on whether or not to install a vibration-based SHM system on a structure, for a wide range of SHM use cases.
		\end{abstract}
		
		
	\end{frontmatter}
	
	
	\section{Introduction}
	
	Operation and maintenance (O\&M) of structures and infrastructures addresses various potential threats (e.g., gradual deterioration, extreme events) that can adversely affect the intended performance of these systems. This creates the need for inspection, maintenance and repair actions throughout a system's life-cycle, which come at a large cost \cite{Sasidharan_2021, Nielsen_2010}. In the current approach to O\&M, visual inspection still remains the primary, and oftentimes sole, source of information on the condition of a structure over its life-cycle.
	
	Structural Health Monitoring (SHM) is defined as a continuous, automated, online process for damage assessment, whose ultimate goal is to provide cradle-to-grave system state awareness \cite{Farrar_Worden_2013}. Continuous vibration-based SHM systems offer a great potential for facilitating and enhancing the O\&M decision making process. Despite comprehensive scientific and practical developments in the field of SHM \cite{Farrar_Worden_2007}, adoption of SHM systems on real-world structures and infrastructure systems falls short of the mark \cite{Cawley_2018}, due to a number of challenges. These include the fact that damage-sensitive features can be very sensitive to changes in environmental and operational conditions (EOCs) \cite{Peeters_2000, Moser_2011}. Furthermore, since there exist only a limited number of deployments of such systems on real-world structures, and since data from damage states are rarely available, it is often not clear how to make efficient use of acquired monitoring data for statistical decision making in a supervised learning mode. Besides, it is difficult to convince owners and operators of the potential economic benefit of installing SHM systems \cite{Ye_2021}. 
	
	A clear need exists for offering actionable use cases, elaborating on the manner in which SHM systems can lead to enhanced management of structural deterioration, and how these can eventually inform optimal maintenance decisions over the structural life-cycle. A wide range of diverse SHM systems is continuously being developed and made available for application on a wide range of structures for performing various damage detection tasks. In designing a case-specific SHM system, one should start by defining the specific target structure and the associated damages that one pursues to detect with this system \cite{Farrar_Worden_2007}. 
	
	Bayesian decision analysis \cite{Raiffa_1961} offers a formal mathematical framework which enables investigating a-priori, i.e., prior to the installation of the SHM system, how monitoring data from a specific SHM system can inform inspection, maintenance or repair actions over the life-cycle of a structure subjected to a certain type of damage. Bayesian decision analysis further enables quantification of the effect of SHM systems on structural life-cycle costs through the Value of Information (VoI) \cite{Pozzi_2011, Straub_2014, Thons_2018, Giordano_2020, Verzobio_2021, Nielsen_2021, Kamariotis_2022}.
	
	The current paper presents a Bayesian decision analysis framework for the quantification of the value of continuous vibration-based SHM, which is applicable across different time scales. In contrast to most works to date, which utilize simple idealized models of the information obtained from monitoring, we employ a realistic model of a vibration-based SHM system, considering modal data that is continuously identified via operational modal analysis (OMA) schemes. Our analysis also includes the effect of environmental variability on the identified modal data, a key issue in vibration-based SHM. Finally, the framework includes the full sequence of inspection and maintenance decisions throughout the structural life-cycle rather than just individual decisions, as in most of the literature.
	
	The paper is organized as follows. Section 2 offers a fundamental classification of SHM use cases in terms of the associated time scales for decision making. Section 3 presents the proposed Bayesian decision analysis framework and the associated Value of SHM (VoSHM) metric. Section 4 introduces an environmental variability model and a structural deterioration model, and discusses sequential Bayesian learning of these models using continuous SHM data and inspection data. Section 5 contains an algorithmic summary of the presented framework for quantifying the VoSHM. Section \ref{sec:Num_inv} demonstrates application of this framework on different SHM use cases with the aid of a numerical model of a deteriorating two-span bridge system. Finally, Section 7 discusses and concludes this work.
	
	\section{SHM use cases across different time scales}
	\label{sec:Chapter_2}
	
\begin{figure}[ht!]
		\centerline{
			\includegraphics[width=0.80\textwidth]{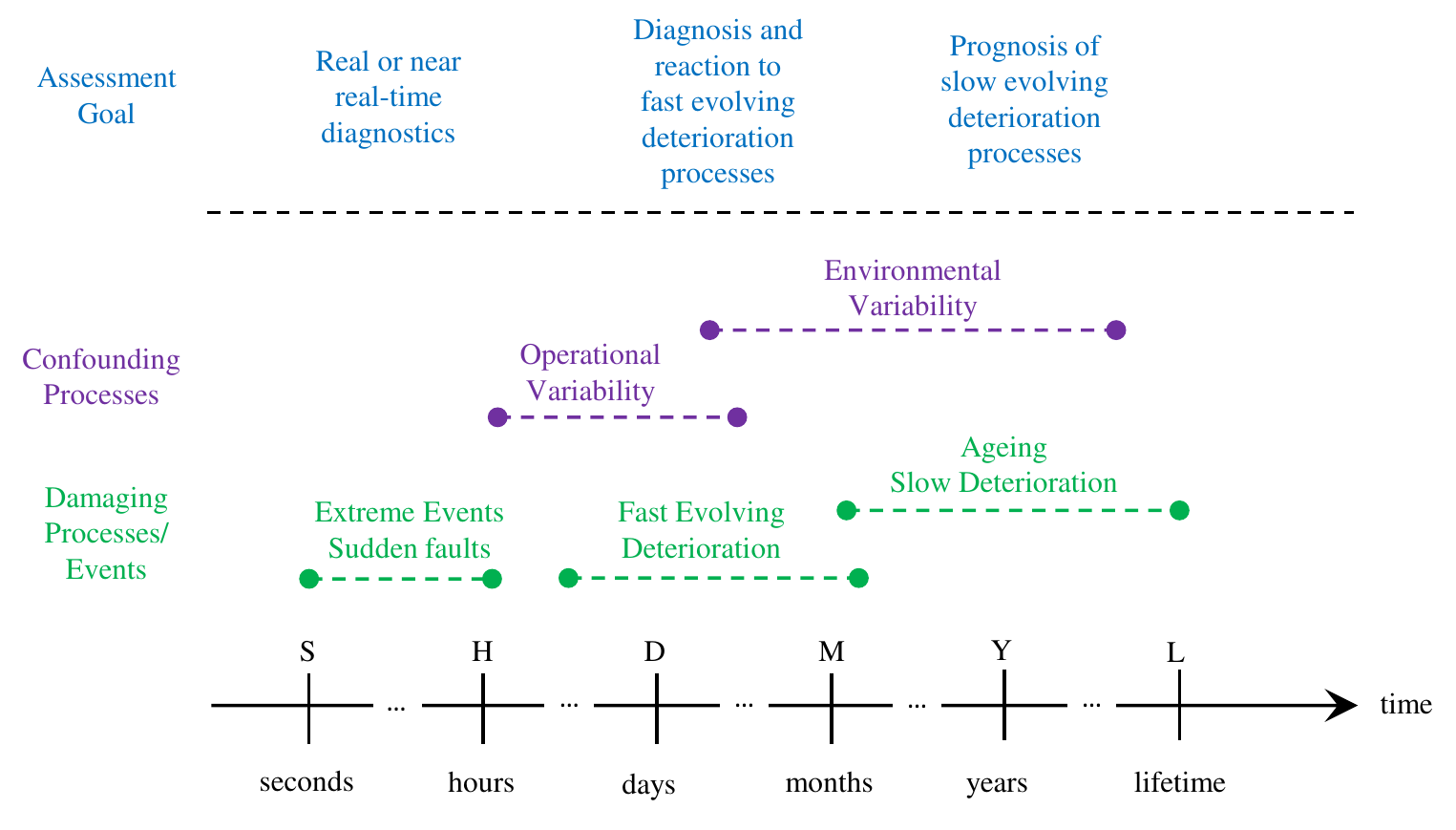}
		}
		\caption{SHM use case-dependent time scales for decision making}
		\label{fig: time_scales}
	\end{figure}

	In this section, we discuss various SHM use cases in relation to the different time scales at which they support decision making for infrastructure (Figure \ref{fig: time_scales}), with a particular focus on application of SHM systems on civil structures. The main challenges associated with an efficient use of the data provided by an SHM system for decision support at each time scale are laid out.
	
	\subsection{Real-time or near real-time diagnostics (seconds to hours)}
	\label{subsec:Near real-time}
	At this time scale, fast, almost online, detection of flaws or abnormalities is sought. When discussing a scale of seconds, one typically refers to real-time tracking and diagnostic tasks. This is particularly relevant in the context of control, where a possible failure of the control system (e.g. active vibration control) should be computed almost instantly. Near real-time tasks relax the requirements on the speed of reaction but still call for accelerated diagnosis, typically linked to emergency operations (e.g., smart tagging of buildings after an earthquake, powering down a wind turbine after a lightning strike, promptly deciding on whether to close down a bridge after a flood occurrence \cite{PREGNOLATO_2017}). SHM can be valuable in informing near real-time decisions for avoiding catastrophic failures (e.g., bridge support or wind turbine blade failure), or avoiding unnecessary closures and down-time after the occurrence of an extreme event (e.g., rapid seismic loss assessment of structures using near real-time data \cite{Tubaldi_2022}). Quantifying the VoSHM over the system lifetime requires a model of the occurrence of extreme events (shocks) \cite{Sanchez-Silva_2011, Jia_2018} that induce these abrupt failures (e.g., wind, flood, earthquake). A main challenge at this time scale stems from the real-time nature of required diagnostics, which implies fast computation, as well as from the masking influence of varying EOCs on the detection capabilities. Moreover, derivation of robust data-driven diagnostics in a fully unsupervised and automated manner is an intricate task. It is difficult to achieve higher-end SHM tasks beyond damage detection, such as damage localization or quantification, in a purely ``online" data-driven fashion, without use of a dedicated model. 
	
	\subsection{Fast-evolving deterioration processes (days to months)}
	
	Here, the objective lies in identifying structural deterioration processes with a rather fast rate of evolution, or capturing sudden damage increments caused by shock deterioration, which could endanger capacity, availability or serviceability. At this time scale, application of SHM should assist in avoiding failure and ensuring the desired safety and performance levels. Examples of damage types induced by effects at this time scale could be, e.g., a shaft failure on a train bogie after crack initiation, concrete bridge failure due to a fast evolving Akali-silica reaction (ASR) process \cite{Jensen_2004}, or freeze-thaw related damages \cite{Sakulich_2012}. Quantification of the VoSHM in such cases requires models for simulating such accelerated or shock deterioration processes \cite{Sanchez-Silva_2011, Jia_2018}. Furthermore, models are needed for the estimation of the structural reliability \cite{Melchers_2017}, which can be used as a metric to evaluate structural performance. Accounting for EOC variability can pose a significant challenge at this time scale as well.
	
	\subsection{Slow-evolving deterioration processes (years/life)}
	Over larger time spans, SHM can be used to support decisions on corrective, preventive or predictive maintenance associated with slowly-evolving gradual deterioration processes, such as fatigue or corrosion. Models for simulating such deterioration processes over the life-cycle \cite{Sanchez-Silva_2011, Jia_2018}, as well as models for estimating the structural reliability and its updating using Bayesian methods \cite{Straub_2020}, are indispensable for a VoSHM analysis. At this time scale, main challenges include EOC variability, loads which are typically increasing over time (e.g. heavier trucks), abrupt changes in the assumed deterioration model (e.g., due to shock events).
	
	\subsection{Summary}
	The operation of real structural systems typically involves a combination of the above-mentioned potential threats. A successful management strategy should prescribe a plan for addressing these threats throughout the structural life-cycle. The main goal of this paper is to show that adoption of the proposed Bayesian decision analysis framework can lead to a comprehensive tool for performing quantitative VoSHM studies across these diverse time scales. Eventually, this framework can act as an a-priori decision support tool for the crucial decision on whether adoption of a specific SHM system can be cost-beneficial.
	
	\section{Bayesian decision analysis framework for the quantification of the VoSHM}
	\label{sec:Framework}
	
	In a decision analysis, the goal is to find the optimal set of actions which maximize the expected utility. For most engineering applications, utility can be equated with the negative total life-cycle costs \cite{Streicher_2004}. Therefore, the problem translates to finding the optimal set of actions over the structural lifetime, that minimize the expected total life-cycle costs. 
	
	The total life-cycle costs $C_{\text{tot}}(\boldsymbol{X}, \boldsymbol{a})$ are defined as a function of a random vector $\boldsymbol{X}$, containing the parameters of the stochastic deterioration model and the structural response quantities, as well as a set of actions $\boldsymbol{a}$ that are performed on the system at different points in time over its life-cycle, such as inspection, repair or maintenance. Different cost components synthesize the overall costs $C_{\text{tot}}(\boldsymbol{X}, \boldsymbol{a})$ \cite{Frangopol_1997}. Because the initial cost of construction and the decommissioning cost at the end of the structure's lifetime are not affected by the SHM, they can be ignored for the VoSHM analysis. Therefore, $C_{\text{tot}}(\boldsymbol{X}, \boldsymbol{a})$ is the sum of the inspection costs $C_{\text{I}}$, the maintenance costs $C_{\text{M}}$, and the risk (the expected cost of failures) $R_{\text{F}}$ over the lifetime of the structure: 
	\begin{linenomath}\begin{equation}
			C_{\text{tot}}(\boldsymbol{X}, \boldsymbol{a}) = C_{\text{I}}(\boldsymbol{X}, \boldsymbol{a}) +
			C_{\text{M}}(\boldsymbol{X}, \boldsymbol{a}) + R_{\text{F}}(\boldsymbol{X}, \boldsymbol{a}) 
			\label{Ctot_decomposition}
	\end{equation}\end{linenomath}
	
	 The goal is to find at any decision time step $t$ the optimal set of inspection, repair and maintenance actions $a_t$ that lead to a balance between the discounted cost of these actions and the failure risk \cite{Straub_2005, Papakonstantinou_2014}. The risk of a failure event $F(t)$ at time step $t$, can be quantified via the outcome of a structural reliability analysis \cite{Straub_2020, Melchers_2017}. 
	
	A prior decision analysis is performed, where one only has access to prior information on the random vector $\boldsymbol{X}$. In a prior decision analysis, the optimal set of actions over the structural life-cycle is found as:
	\begin{linenomath}\begin{equation}
			\boldsymbol{a}_{opt} = \underset{\boldsymbol{a}}{\argmin}\boldsymbol{\text{E}}_{\boldsymbol{X}}[C_{\text{tot}}(\boldsymbol{X}, \boldsymbol{a})]
			\label{Prior_decision}
	\end{equation}\end{linenomath}
	where $\boldsymbol{\text{E}}_{\boldsymbol{X}}$ is the expectation with respect to the prior distribution $f_{\boldsymbol{X}}(\boldsymbol{x})$.
	
	The benefit of monitoring is that it can provide information that reduces the uncertainty on the deterioration state and model parameters, thus enabling monitoring-informed risk estimation, which ultimately leads to better decisions. Once monitoring data $\boldsymbol{z}$ becomes available, one can perform a posterior decision analysis, in order to find the set of actions that are optimal conditional on the monitoring data. This requires solving the following optimization problem: 
	\begin{linenomath}\begin{equation}
			\boldsymbol{a}_{opt \mid \boldsymbol{z}}  = \underset{\boldsymbol{a}}{\argmin}\boldsymbol{\text{E}}_{\boldsymbol{X}\mid \boldsymbol{z}}[C_{\text{tot}}(\boldsymbol{X}, \boldsymbol{a})]
			\label{Posterior_decision}
	\end{equation}\end{linenomath}
	where the expectation $\boldsymbol{\text{E}}_{\boldsymbol{X} \mid \boldsymbol{z}}$ operates over the distribution of $\boldsymbol{X}$ conditional on $\boldsymbol{z}$. Therefore, prior to solving the optimization problem of Equation \eqref{Posterior_decision}, one has to perform Bayesian analysis \cite{Gelman_2013, Sarkka_2013} to estimate the posterior distribution $f_{\boldsymbol{X} \mid \boldsymbol{z}}(\boldsymbol{x}\mid \boldsymbol{z})$, which is presented in Section \ref{sec:BA}.
	
	The monitoring data $\boldsymbol{z}$ becomes available only after installation of an SHM system. However, typically one is interested in investigating the potential benefit of the deployment of a specific SHM system prior to its installation. In that case, a dedicated model of the SHM system must be employed to allow for probabilistic predictions of the monitoring information $\boldsymbol{Z}$ that one expects to extract with monitoring, for given sampled realizations of the random vector $\boldsymbol{X}$. Upper case $\boldsymbol{Z}$ therefore denotes the yet unknown monitoring information, while a realization of $\boldsymbol{Z}$ is denoted by lower case $\boldsymbol{z}$. The value-of-information analysis is conducted prior to observing any actual monitoring data $\boldsymbol{z}$, relying instead on simulated data. Hence this is known as a preposterior analysis \cite{Raiffa_1961}. The expected total life-cycle cost in such a preposterior decision analysis is written as:
	\begin{linenomath}\begin{equation}
			\boldsymbol{\text{E}}_{\boldsymbol{X}, \boldsymbol{Z}}[C_{\text{tot}}(\boldsymbol{X}, \boldsymbol{a}_{opt \mid \boldsymbol{z}})]
			\label{Preposterior_cost}
	\end{equation}\end{linenomath}
	where one can observe that the expectation $\boldsymbol{\text{E}}_{\boldsymbol{X}, \boldsymbol{Z}}$ jointly operates over $\boldsymbol{X}$ and $\boldsymbol{Z}$. Using a sampling-based approach, for a realization of the structural system's uncertain parameters and response state space, i.e., for a given $\boldsymbol{X}$, one has to employ a model of the investigated SHM system to probabilistically predict the monitoring data $\boldsymbol{z}$ (observations) that the SHM system will provide over the life-cycle. With such a model, one can sample one (or more) realizations of $\boldsymbol{z}$. Any type of information that can be extracted from the SHM system and that can be used as a damage sensitive feature can be considered as a sampled $\boldsymbol{z}$ realization (see Section \ref{subsec:SHM_sampling}). For each sampled $\boldsymbol{z}$ realization, one subsequently needs to perform a posterior decision analysis, as in Equation \eqref{Posterior_decision}, to find the posterior optimal set of actions $\boldsymbol{a}_{opt \mid \boldsymbol{z}}$. This analysis has to be performed multiple times, for a sufficiently large set of samples of $\boldsymbol{X}, \boldsymbol{Z}$.
	The Value of Information (VoI) is quantified by taking the difference between the expected total life-cycle cost in the prior decision analysis and the expected total life-cycle cost in the preposterior decision analysis, as follows:
	\begin{linenomath}\begin{equation}
			VoI = \boldsymbol{\text{E}}_{\boldsymbol{X}}[C_{\text{tot}}(\boldsymbol{X}, \boldsymbol{a}_{opt})]-
			\boldsymbol{\text{E}}_{\boldsymbol{X}, \boldsymbol{Z}}[C_{\text{tot}}(\boldsymbol{X}, \boldsymbol{a}_{opt \mid \boldsymbol{z}})]
			\label{VoI}
	\end{equation}\end{linenomath}
	
	A VoI analysis offers a formal framework for quantifying the effect of SHM systems on structural life-cycle costs \cite{Pozzi_2011, Straub_2014, Straub_2017}. However, quantifying the VoI in this way is not very informative for system owners and operators, as it assumes that in the reference case neither inspection nor monitoring data will be available. It is seldom the case that no inspection or monitoring takes place throughout the whole life-cycle of a structure. Typically, intermittent visual inspection schemes are adopted by operators, with targeted non-destructive evaluations also complementing inspection when needed \cite{Cawley_2018}. Therefore, to demonstrate the potential benefit of deploying continuous SHM systems on structures, as compared against the typical case of intermittent visual inspections, a more specialized metric, the Value of Structural Health Monitoring (VoSHM)\cite{Andriotis_2021}, can be introduced:	\begin{linenomath}\begin{equation}
			VoSHM = \boldsymbol{\text{E}}_{\boldsymbol{X}, \boldsymbol{Z}_{insp}}[C_{\text{tot}}(\boldsymbol{X}, \boldsymbol{a}_{opt \mid \boldsymbol{z}_{insp}})]-
			\boldsymbol{\text{E}}_{\boldsymbol{X}, \boldsymbol{Z}_{insp}, \boldsymbol{Z}_{SHM}}[C_{\text{tot}}(\boldsymbol{X}, \boldsymbol{a}_{opt \mid\boldsymbol{z}_{insp}, \boldsymbol{z}_{SHM}})]
			\label{VoSHM}
	\end{equation}\end{linenomath}
	
\begin{figure}
		\centerline{
			\includegraphics[width=\textwidth]{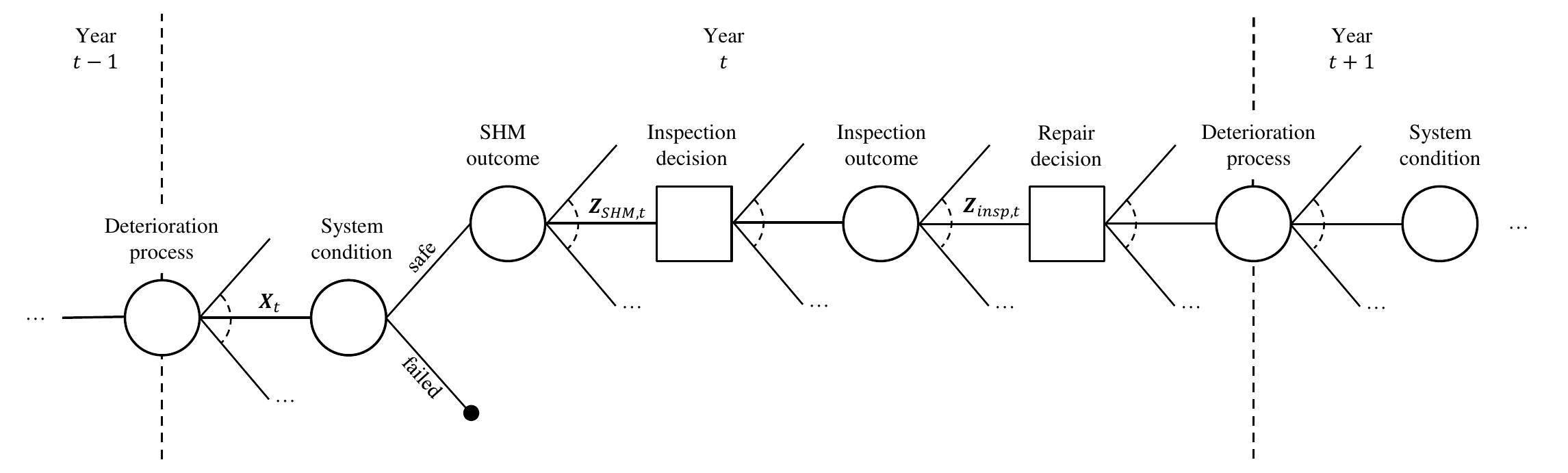}
		}
		\caption{Decision tree illustrating a preposterior decision analysis for a deteriorating structure, which is continuously monitored from an SHM system, which can additionally be inspected and repaired.}
		\label{f:decision_tree}
	\end{figure}

	The formulation of Equation \eqref{VoSHM} implies that the quantification of the VoSHM emerges from the solution of two different preposterior decision analyses. In Equation \eqref{VoSHM}, we no longer use the generic variable $\boldsymbol{Z}$ to denote observations at large, but instead enforce a distinction between the data obtained from the continuous SHM system, denoted as $\boldsymbol{Z}_{SHM}$, and the data obtained via intermittent visual inspections, denoted as $\boldsymbol{Z}_{insp}$. Similarly to the requirement of a model of the monitoring system to probabilistically predict $\boldsymbol{Z}_{SHM}$, one further needs a model that enables probabilistic predictions of the inspection data $\boldsymbol{Z}_{insp}$. In $\boldsymbol{\text{E}}_{\boldsymbol{X}, \boldsymbol{Z}_{insp}}[C_{\text{tot}}(\boldsymbol{X}, \boldsymbol{a}_{opt \mid \boldsymbol{z}_{insp}})]$ the total expected life-cycle cost is computed for the case when only inspection data is available. In $\boldsymbol{\text{E}}_{\boldsymbol{X}, \boldsymbol{Z}_{insp}, \boldsymbol{Z}_{SHM}}[C_{\text{tot}}(\boldsymbol{X}, \boldsymbol{a}_{opt \mid\boldsymbol{z}_{insp}, \boldsymbol{z}_{SHM}})]$, the total expected life-cycle cost is computed for the case of continuous monitoring data, enriched by additional inspection information. The latter problem is illustrated by the decision tree in Figure \ref{f:decision_tree}. The defined VoSHM metric assumes that an SHM system will be used in conjunction with some additional inspection policy (which will be informed by the SHM outcome), as -in a practical setting- operators would not allocate sufficient trust on a completely autonomous and unsupervised monitoring system, thus entirely replacing inspections.
	
	\subsection{Solution of the sequential decision problem via adoption of heuristics}
	\label{subsec:heuristics}
	
	A key challenge of a preposterior decision analysis is the identification of the optimal set of actions conditional on data $\boldsymbol{a}_{opt \mid \boldsymbol{z}}$. The optimization of inspection and maintenance plans forms a stochastic sequential decision problem \cite{Luque_2019, Kochenderfer_2015}, the solution of which requires large computational efforts \cite{Straub_2014, Morato_2022}. Numerous algorithms are available for the solution of this problem, e.g., via proposal of a set of simple decision heuristics \cite{Luque_2019,Bismut_2021}, through partially observable Markov decision processes (POMDPs) \cite{Andriotis_2021, Morato_2022}, or through deep reinforcement learning \cite{Andriotis_2019}.
	
	The concepts of policies and strategies have been introduced for the solution of sequential decision problems \cite{Jensen_2007, Bismut_2021}. A policy at time $t$ is a set of rules which prescribes the decisions to be made at time $t$, based on all the structural state information available up to that time, i.e., past inspection and monitoring data $\boldsymbol{Z}$ and performed actions. Here, a policy at time $t$ answers the following two questions: 1) `Inspect the structure?', 2) `Repair the structure?'. A strategy $S$ is the set of policies for all time steps $t$ of the decision time horizon, which typically reflects the intended lifetime of the structure. If a strategy consists of policies which are the same at all times $t$, then the strategy is called stationary.
	
	Heuristics, which are simple and intuitive rules that are easily understood by engineers and operators, can be used to parametrize a stationary strategy \cite{Luque_2019, Bismut_2021}. The three heuristics applied herein are described below. These are based on the premise that structural performance at any time $t$ conditional on inspection and monitoring results can be assessed via the structural reliability estimate. With Bayesian methods, past inspection and monitoring information can be used to update the structural reliability estimate \cite{Straub_2020}. 
	\begin{enumerate}
		\item \emph{Reliability threshold for inspections $p^{I}_{th}$.} An inspection is performed at any time step before the updated estimate of the structural reliability exceeds $p^{I}_{th}$.
		\item \emph{Fixed-interval periodic inspections $\Delta t_I$.} Periodic inspections are performed every $\Delta t_I$ years.
		\item \emph{Reliability threshold for repair $p^{R}_{th}$.} A repair is performed at any time step before the updated estimate of the structural reliability exceeds $p^{R}_{th}$.
	\end{enumerate}
	
	We define the heuristic parameter vector $\boldsymbol{w} = [p^{I}_{th}, \Delta t_I, p^{R}_{th}]$. A strategy parametrized by the heuristic parameter vector $\boldsymbol{w}$ is denoted $S_{\boldsymbol{w}}$. With the use of heuristics, the optimal set of actions conditional on data $\boldsymbol{z}$, $\boldsymbol{a}_{opt \mid \boldsymbol{z}}$, is approximated by the applied optimal heuristic parameter values, which drastically reduces the space of solutions to the decision problem. In this way, the solution of the sequential decision problem boils down to finding the optimal set of heuristic parameter values $\boldsymbol{w}^*$ for which the strategy $S_{\boldsymbol{w^*}}$ is optimal, i.e., leads to the minimum expected total life-cycle costs.
	\begin{linenomath}\begin{equation}
			S_{\boldsymbol{w}^*} = \underset{\boldsymbol{w}}{\argmin}\boldsymbol{\text{E}}\left[C_{\text{tot}}\left(\boldsymbol{X}, \boldsymbol{Z}, \boldsymbol{w}\right)\right]
			\label{Optimal_heuristic_strategy}
	\end{equation}\end{linenomath}
	
	With the use of heuristics, Equation \eqref{VoSHM} can be reformulated as follows:
	\begin{linenomath}\begin{equation}
			VoSHM =     \boldsymbol{\text{E}}_{\boldsymbol{X}, \boldsymbol{Z}_{insp}}\left[C_{\text{tot}}\left(\boldsymbol{X}, \boldsymbol{Z}_{insp}, \boldsymbol{w}_1^*\right)\right]-
			\boldsymbol{\text{E}}_{\boldsymbol{X}, \boldsymbol{Z}_{insp}, \boldsymbol{Z}_{SHM}}\left[C_{\text{tot}}\left(\boldsymbol{X}, \boldsymbol{Z}_{insp}, \boldsymbol{Z}_{SHM}, \boldsymbol{w}_2^*\right)\right]
			\label{VoSHM_heuristic}
	\end{equation}\end{linenomath}
	Equation \eqref{VoSHM_heuristic} indicates that the optimal heuristic parameter vectors $\boldsymbol{w}_1^*$ and $\boldsymbol{w}_2^*$ emerging from the solution of the two different preposterior decision analyses will differ. Obtaining $\boldsymbol{w}_2^*$ would typically require much larger computational effort than determining $\boldsymbol{w}_1^*$. 
	
	The problem can be further simplified by replacing the optimization of the heuristic parameters in Equation \eqref{Optimal_heuristic_strategy} by a choice based on expert assessment, which better reflects what is typically done in practice, where optimization is rarely performed.
	
	A Monte Carlo approach can be used to estimate the expected total life-cycle cost for given heuristic strategies for both terms in Equation \eqref{VoSHM_heuristic}. The Monte Carlo approximation of the $VoSHM$ is
	\begin{linenomath}\begin{equation}
			VoSHM \approx \frac{1}{n_{\text{MCS}}} \sum_{i=1}^{n_{\text{MCS}}}\left[C_{\text{tot}}\left(\boldsymbol{x}^{(i)},\boldsymbol{z}_{insp}^{(i)}, \boldsymbol{w}_1^*\right) - C_{\text{tot}}\left(\boldsymbol{x}^{(i)}, \boldsymbol{z}_{insp}^{(i)}, \boldsymbol{z}_{SHM}^{(i)}, \boldsymbol{w}_2^*\right)\right]
			\label{VoSHM_heuristic_MCS}
	\end{equation}\end{linenomath}
	wherein $\boldsymbol{x}^{(i)}$ are random samples drawn from the prior distribution $f_{\boldsymbol{X}}(\boldsymbol{x})$, $\boldsymbol{z}_{insp}^{(i)}$ are samples from the inspection likelihood function $f_{\boldsymbol{Z}_{insp}|\boldsymbol{X}}(\cdot|\boldsymbol{x}^{(i)})$ and $\boldsymbol{z}_{SHM}^{(i)}$ are sampled data from the model of the SHM system.
	
	\subsubsection{Cost breakdown}
	For a given heuristic strategy, i.e., for a given $\boldsymbol{w}$, and for given inspection $\boldsymbol{z}_{insp}^{(i)}$ and monitoring outcomes $\boldsymbol{z}_{SHM}^{(i)}$, the inspection and repair actions over the structural life-cycle are fixed, and the costs are computed as follows:
	\begin{linenomath}\begin{equation}
			C_{\text{I}}\left(\boldsymbol{x}^{(i)}, \boldsymbol{z}_{insp}^{(i)}, \boldsymbol{z}_{SHM}^{(i)}, \boldsymbol{w}\right) = \sum_{j=1}^{n_{\text{insp}}}\gamma\left(t_{insp}^{(j)}\right)\widetilde{c_{\text{I}}}
			\label{C_insp}
	\end{equation}\end{linenomath}
	\begin{linenomath}\begin{equation}
			C_{\text{R}}\left(\boldsymbol{x}^{(i)}, \boldsymbol{z}_{insp}^{(i)}, \boldsymbol{z}_{SHM}^{(i)}, \boldsymbol{w}\right) = \sum_{j=1}^{n_{\text{rep}}}\gamma\left(t_{rep}^{(j)}\right)\widetilde{c_{\text{R}}}
			\label{C_rep}
	\end{equation}\end{linenomath}
	where $\widetilde{c_{\text{I}}}$, $\widetilde{c_{\text{R}}}$ are the costs of an individual inspection and repair respectively, and $\gamma(t) = \frac{1}{(1+r)^t}$ is the discounting function, with $r$ denoting the annually compounded discount rate.
	
	The failure risk $R_F$ is defined as:
	\begin{linenomath}\begin{equation}
			R_{\text{F}}\left(\boldsymbol{x}^{(i)}, \boldsymbol{z}_{insp}^{(i)}, \boldsymbol{z}_{SHM}^{(i)}, \boldsymbol{w}\right)= \sum_{j=1}^{T}\gamma(t_j)\widetilde{c_{\text{F}}} \left\{ \text{Pr}[F(t_j)|\boldsymbol{x}^{(i)}] - \text{Pr}[F(t_{j-1})|\boldsymbol{x}^{(i)}]\right\}
			\label{Risk}
	\end{equation}\end{linenomath}
	where $\widetilde{c_{\text{F}}}$ is the cost of a failure event, and $Pr[F(t_j)|\boldsymbol{x}^{(i)}]$ is the conditional probability of failure of the structure up to time $t_j$, conditional on a sampled value $\boldsymbol{x}^{(i)}$ of the random vector $\boldsymbol{X}$. Its computation forms a structural reliability problem, which is laid out in Section 4.3.
	
	\subsection{Summary of the framework for quantifying the value of vibration-based SHM}
	
	Section \ref{sec:Framework} presents the Bayesian decision analysis framework in a general way, and then specifically introduces the VoSHM metric in Equation \eqref{VoSHM}, which forms the basis for the presented framework for the quantification of the value of vibration-based SHM.
	
	Equation \eqref{VoSHM} reveals that in order to perform a VoSHM analysis, one needs to specify different models and computational approaches. A probabilistic model of the random vector $\boldsymbol{X}$ needs to be defined, which in this paper is detailed in Section \ref{sec:BA}, where it is exemplified for a specific environmental variability model and deterioration model. A case-specific model of the vibration-based SHM system must be defined to allow for probabilistic predictions of the SHM outcomes $\boldsymbol{Z}_{SHM}$ (see Figure \ref{fig:SHM_sampling}), which can be made specific only in accordance with the case study at hand, while, lastly, a specific inspection model is required for sampling inspection outcomes $\boldsymbol{Z}_{insp}$ (see Equation \eqref{insp_likelihood_function}). One further needs to define the stochastic sequential decision-making problem for the optimization of inspection and maintenance plans and choose the corresponding computational method for its solution. The solution to the latter problem using heuristics has been introduced in Subsection \ref{subsec:heuristics}. Section \ref{sec:Summary} summarizes the algorithm for evaluation of Equation \eqref{VoSHM}, i.e., it provides the computational specifics of the framework.
	
	\section{Environmental variability modeling, deterioration modeling and Bayesian analysis}
	\label{sec:BA}
	
	\subsection{Environmental variability model}
	The premise of vibration-based SHM methods is that damage induces changes in the structural system's modal characteristics, e.g., the system's natural frequencies and mode shapes \cite{Farrar_Worden_2013}. Thereby it must be considered that varying environmental and operational conditions also affect the system's modal characteristics. Temperature affects the stiffness (the effective Young's modulus) of civil structures \cite{Peeters_2000, Moser_2011, Martin_Sanz_2020}. The resulting changes in the system's modal characteristics owing to temperature variability can often be more prominent than the changes due to significant damage. Related to the effect of the temperature variability on the VoSHM, the authors have shown in \cite{Kamariotis_2022_c} that not properly accounting for the environmental variability present in the SHM data can have a detrimental effect on the maintenance decisions triggered by the SHM system. Accounting for the effects of temperature variability is therefore of utmost importance within a vibration-based damage identification framework, and various ways to do that have been suggested in literature \cite{Moser_2011, Figueiredo_2011, Spiridonakos_2016}. 
	
	The dependence of the system's identified natural frequencies on temperature may often be nonlinear \cite{Moser_2011}, especially in environments where the structure can experience below-freezing temperatures, and this nonlinearity has to be taken into account in the modeling. The following model for the Young's modulus as a function of temperature is employed, which was previously used to capture the dependence present in real data obtained from a bridge structure \cite{Behmanesh_2016}.
	\begin{linenomath}\begin{equation}
			E(T_t) = \theta(T_t) \cdot E_0
			\label{E_eff}
	\end{equation}\end{linenomath}
	\begin{linenomath}\begin{equation}
			\theta(T_t) = Q + H \cdot T_t + U \cdot \left(1- \erf\left(\frac{T_t-Y}{\tau}\right)\right)
			\label{theta_temp}
	\end{equation}\end{linenomath}
	$E_0$ is the nominal value for the Young's modulus at a reference temperature of $20^{\circ}$C. The effective Young's modulus at a time instance $t$, for a given temperature $T_t$, is given by Equation \eqref{E_eff}. The structural parameter $\theta$ is introduced as a modification factor for the effective Young's modulus at a given temperature $T_t$. $\theta(T_t)$ is a stochastic function of temperature, shown in Equation \eqref{theta_temp}, which is described by a parameter vector $[Q, H, U, Y, \tau]$ of independent random variables, each following an assigned prior probability distribution. $Q$ models the intercept of the linear trend in the above-freezing temperature range, while $H$ models the slope of the linear trends observed in both above and below-freezing temperatures, after the nonlinear transition around temperature $Y$. $\tau$ models the transition range, while $U$ defines the size of increase in the Young's modulus at the end of the transition to the below-freezing temperature range.
	
	\subsubsection{Bayesian learning of the environmental variability model}
	\label{subsubsec:Bayesian_environmental}
	
	Let us consider a case when an SHM system is installed on a civil structure from the beginning of its operation. Typically, vibration-based SHM systems rely on the deployment of acceleration sensors, which can provide continuous dynamic response measurements in the form of acceleration time series, for unknown ambient excitation. These can be subsequently processed by output-only operational modal analysis (OMA) schemes, e.g., the stochastic subspace identification (SSI) algorithm \cite{Peeters_1999}, for identifying the system's modal characteristics. Temperature sensors can further be easily and inexpensively deployed on a structure, which can provide ambient temperature measurements. With the assumption that the structure will be in a healthy state and that no damage will be present at the beginning of its service life, one can make use of modal data identified at different temperatures in the first few months of operation (half a year to one year will usually be needed to get `extreme' temperatures at both ends of the scale), in order to learn the underlying dependence of the Young's modulus on temperature (E-T) via Bayesian analysis. 
	
	The data obtained from such an SHM system, comprising acceleration and temperature sensors, can be summarized as different sets $\{ \widetilde{\lambda}_{t_m}, \widetilde{T}_t; m = 1,.., N_m$\} of vectors of the $N_m$ lower system eigenvalues $\widetilde{\lambda}_{t_m}$ identified via an OMA procedure at time $t$ for a temperature $\widetilde{T}_t$. The modal eigenvalues are $\widetilde{\lambda}_{t_m} = (2\pi \widetilde{f}_{t_m})^2$, where $\widetilde{f}_{t_m}$ are the modal eigenfrequencies.
	
	Consider a linear finite element (FE) model $\mathcal{G}$ of the structural system, parametrized via Equation \eqref{E_eff}, which is used to predict the modal eigenvalues for different input values of the effective Young's modulus. The goal of the Bayesian inverse problem is to estimate the parameters $[Q, H, U, Y, \tau]$ of the stochastic model of Equation \eqref{theta_temp}, and their uncertainty, such that the FE model predicted modal eigenvalues $\left\{\lambda_{t_m}^\mathcal{G}\left(E=\theta(\widetilde{T}_t)\cdot E_0\right), m = 1,..,N_m\right\}$ best match the corresponding OMA-identified modal eigenvalues. The joint posterior probability distribution of the updating parameters is obtained via Bayesian analysis.

	The likelihood function can be formulated by assuming a probabilistic model for the discrepancy between the OMA-identified and the FE model predicted modal eigenvalues. The commonly assumed probabilistic model for this discrepancy is a zero-mean Gaussian random variable with standard deviation proportional to the identified eigenvalues. Assuming statistical independence among the $N_m$ identified modes and among the $N_t$ identified modal data sets obtained at different time instances, the likelihood function can be written
	\begin{linenomath}\begin{equation}
			L\left(\widetilde{\lambda}_{t_m}, \widetilde{T}_t; Q, H, U, Y, \tau\right) = \prod_{t=1}^{N_t}\prod_{m=1}^{N_m}N\left(\widetilde{\lambda}_{t_m}- \lambda_{t_m}^\mathcal{G}\left(E=\theta(\widetilde{T}_t)\cdot E_0\right); 0, c_{\lambda m}^2 \widetilde{\lambda}_{t_m}^2\right),
			\label{time_likelihood}
	\end{equation}\end{linenomath}	
	where $N(\,.\,; 0, \sigma^2)$ denotes the value of the normal probability density function with mean zero and variance $\sigma^2$ at a specified location. The factor $c_{\lambda m}$ can be regarded as an assigned coefficient of variation, and its chosen value reflects the total prediction error \cite{Simoen_2015}, accounting for measurement and model uncertainty. For the numerical investigations of Section \ref{sec:Num_inv}, $c_{\lambda m}$=0.02 \cite{Simoen_2015, Kamariotis_2022}. One should be aware that the assumption of independence in Equation \eqref{time_likelihood} typically does not hold. This could be addressed by a hierarchical modeling of the vector $[Q, H, U, Y, \tau]$ \cite{Behmanesh_2015}.
	
	Once a certain number of $N_t$ OMA-identified eigenvalue sets, identified at different temperatures in the initial undamaged state, becomes available, Bayesian analysis can be performed to estimate the posterior distribution of the environmental variability model parameters. One can then input the estimated posterior mean parameter values (denoted by $\mu^{''}$) in Equation \eqref{theta_temp} and obtain a monitoring-informed estimate of the modification factor $\theta^{''}$ as a function of temperature
	\begin{linenomath}\begin{equation}
			\theta^{''}(T_t) = \mu^{''}_{Q} + \mu^{''}_{H} \cdot T_t + \mu^{''}_{U} \cdot \left(1- \erf\left(\frac{T_t-\mu^{''}_{Y}}{\mu^{''}_{\tau}}\right)\right)
			\label{post_theta_temp}
	\end{equation}\end{linenomath}
	Within subsequent damage identification studies, an effective Young's modulus of $E(\widetilde{T_t}) = \theta^{''}(\widetilde{T_t}) \cdot E_0$ is used for a measured temperature $\widetilde{T_t}$.
	
	\subsection{Deterioration model}
	
	Bayesian decision analysis relies on stochastic deterioration models for sampling many different potential ``what-if" scenarios of the damage evolution over the life-cycle of a structure. Furthermore, a VoSHM analysis requires a model of the SHM system, which can provide probabilistic predictions of the life-cycle monitoring data for the different deterioration and temperature samples (see Figure \ref{fig:SHM_sampling}), as well as a model describing the uncertain outcome of visual inspections.
	
	Structural deterioration can be classified in two main categories, 1) gradual deterioration (e.g., due to fatigue, corrosion, crack growth) and 2) shock deterioration (e.g. sudden damages due to extreme events such as earthquakes, floods, etc.), as described in Section \ref{sec:Chapter_2}. Herein, we employ an empirical deterioration model, which is a superposition of a gradual deterioration process and a shock deterioration process. The deterioration state at global time $t$ (from the beginning of the structural lifetime) is described by the following equation: 
	\begin{linenomath}\begin{equation}
			X(t) = At^B\exp{\left(\omega(t)\right)} + \sum_{i=1}^{N(t)}D_i
			\label{deterioration_model}
	\end{equation}\end{linenomath}
	The first part of Equation \eqref{deterioration_model} is a simple rate equation, which models the gradual deterioration process \cite{Elingwood_2005}. Random variable $A$ models the deterioration rate, random variable $B$ models the nonlinearity effect in terms of a power law in time, and $\omega(t)$ models a Gaussian stochastic process noise. The second part of Equation \eqref{deterioration_model} describes a homogeneous compound Poisson process (CPP) \cite{VanNoortwijk_2009, Sanchez_Silva_2016}, and is used to model deterioration due to sporadic shocks, which is typical of, e.g., earthquakes, floods. CPP models incorporate two types of randomness: i) random times of arrival of sporadic shock occurrences and ii) random damage increase due to an occurring shock.  A CPP is a continuous-time stochastic process $\{V(t), t \geq 0 \}$ of the form $V(t) = \sum_{i=1}^{N(t)} D_i $, where:
	\begin{enumerate}
		\item the number of jumps $\{N(t), t \geq 0 \}$ is a Poisson process with rate $\lambda$,
		\item the jumps $\{D_i, i=1,.., N(t)\}$ are independent and identically distributed random variables following a specified probability distribution,
		\item the process $\{N(t), t \geq 0 \}$ and the damage increments $\{D_i, i=1,.., N(t)\}$ are independent.
	\end{enumerate}
	
\begin{figure}
		\centerline{
			\includegraphics[width=0.4\textwidth]{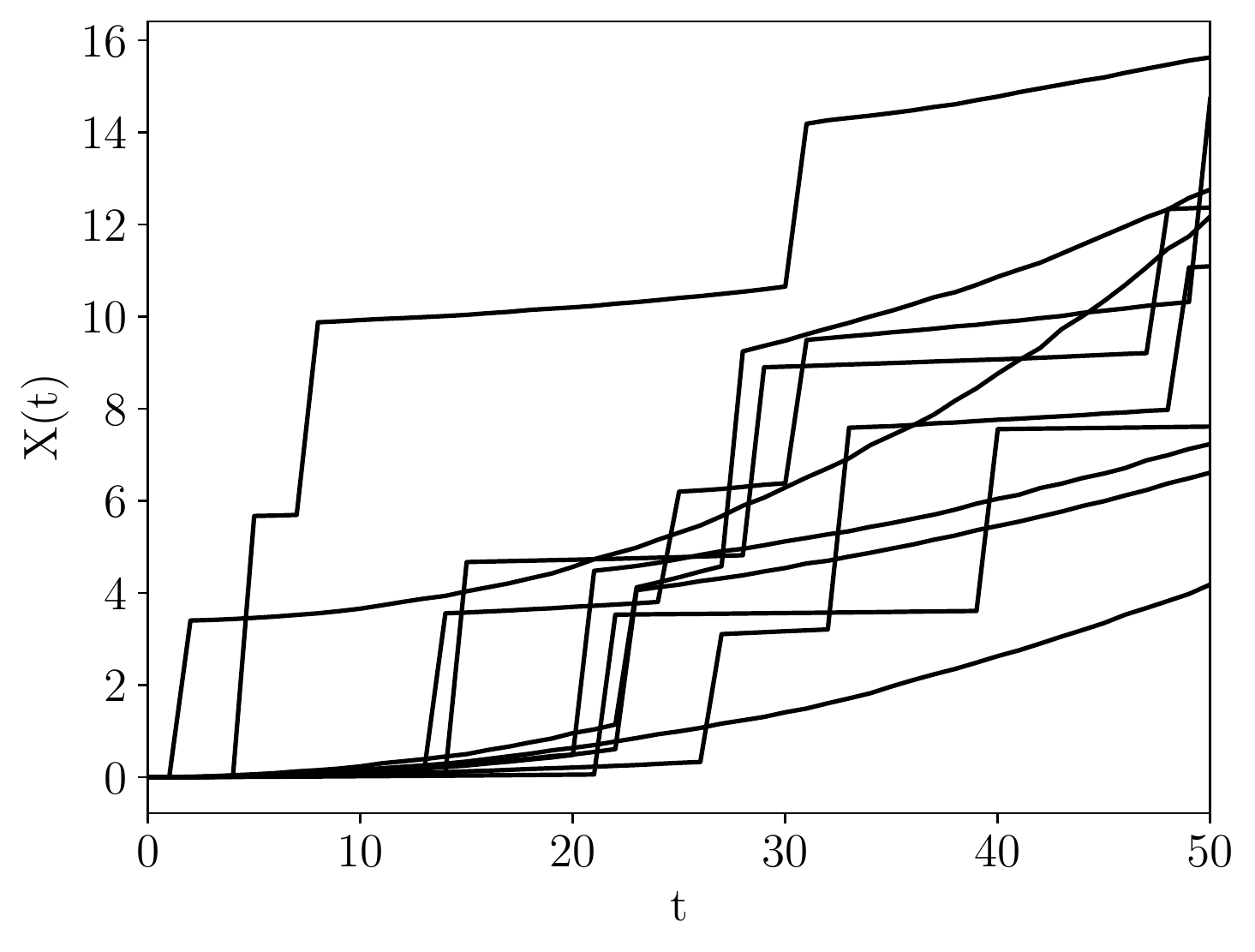}
		}
		\caption{Random sampling from deterioration model of Equation \eqref{deterioration_model} using the parameters of Table \ref{table:Parameters}}
		\label{fig: deterioration model}
	\end{figure}

	The model of Equation \eqref{deterioration_model} assumes that the gradual and shock deterioration processes are independent of each other. Furthermore, it is assumed that the shock deterioration magnitude is independent of the state of the system at the time of the shock event. Ten random realizations of this model are shown in Figure \ref{fig: deterioration model}.
	
	The empirical deterioration model presented above is fairly general and flexible enough to capture a number of the challenges related to performing a VoSHM analysis at different time scales, as described in Section \ref{sec:Chapter_2}. The flexibility of this model is demonstrated in the numerical investigations for different SHM use cases presented in Section \ref{sec:Num_inv}. 
	
	\subsubsection{Sequential Bayesian estimation of deterioration state and parameters}
	The goal is to establish a model that uses the OMA-identified system eigenvalue data, obtained at different points in time and at different temperatures, to jointly update the distribution of the structural deterioration state $X(t)$ and the time-invariant gradual deterioration model parameters $A, B$ of the model in Equation \eqref{deterioration_model}. A discrete-time state-space model is defined, suitable for application of Bayesian filters for monitoring the deterioration. The state space is augmented to include the time-invariant random variables $A$ and $B$ in the estimation. 
	\begin{linenomath}  \begin{align}
			\widetilde{\boldsymbol{X}} &= \begin{bmatrix}
				X(t) \\
				A \\
				B
			\end{bmatrix}
	\end{align}\end{linenomath}	
	The continuous model of Equation \eqref{deterioration_model} is reformulated into a recursive process equation using a central finite difference scheme.
	\begin{linenomath}\begin{equation}
			X_k = X_{k-1} + A_{k-1}B_{k-1}\left(\frac{t_{k-1}+t_k}{2}\right)^{B_{k-1}-1}\Delta t \cdot \text{exp}(\omega_k) + \Delta D_k
			\label{discrete_deterioration_model}
	\end{equation}\end{linenomath}
	where $k$ corresponds to the discrete time instance $t_k$. Time is discretized in yearly intervals $k=1,...,T$, where the k-th interval represents $t \in (t_{k-1}, t_k]$.
	$\Delta D_k$ is the distribution of the CPP jump increments within a time interval $\Delta t$, given by the following cumulative distribution function (CDF)
	\begin{linenomath}\begin{equation}
			F_{\Delta D_k}(d) = \text{exp}(-\lambda \Delta t) + \sum_{i=1}^{\infty} \frac{(\lambda \Delta t)^i}{i!}\text{exp}(-\lambda \Delta t) \cdot F_{\sum_{j=1}^{i}D_j}(d)
			\label{jump_CDF}
	\end{equation}\end{linenomath}
	where $F_{\sum_{j=1}^{i}D_j}(d)$ is the CDF of the $i$-fold convolution of the distribution of $D$ with itself. 
	
	A CPP realization can occur at any point in time $t_{cpp}$. For certain decision cases, one can introduce new decision time intervals in order to incorporate decision making at time $t_{cpp}$, when an extreme event occurs. In such cases, the total number of intervals $n_{int}$ is no longer equal to $T$, i.e, $k=1,...,n_{int}$.
	
	For the time-invariant deterioration model parameters the process equation is as follows:
	\begin{linenomath}\begin{equation}
			[A_{k}, B_{k}]^\intercal = [A_{k-1}, B_{k-1}]^\intercal
			\label{time_inv_process}
	\end{equation}\end{linenomath}
	
	The measurement equation that links the OMA-identified modal eigenvalues with the unknown true deterioration state $X_k$ at time instance $t_k$ is given by 
	\begin{linenomath}\begin{equation}
			\boldsymbol{Z}_{SHM,k} = \boldsymbol{Z}_{k}^\mathcal{G}\left(X_k, E = \theta^{''}(\widetilde{T}_k)E_0\right) + \boldsymbol{\eta}_k
			\label{process_equation}
	\end{equation}\end{linenomath}
	$\boldsymbol{Z}_{SHM,k}$ is the vector of OMA-identified eigenvalues $\{\widetilde{\lambda}_{k_m}, m = 1, ..., N_m\}$ and $\boldsymbol{Z}_{k}^\mathcal{G}$ is the vector of $N_m$ FE model predicted eigenvalues at time $t_k$ and temperature $\widetilde{T}_k$. $\mathcal{G}$ is now parametrized by the deterioration state $X_k$ and the effective Young's modulus. Note that for the effective Young's modulus, the monitoring-informed modification factor $\theta^{''}$ of Equation \eqref{post_theta_temp} is used. To accelerate computations, a surrogate model of $\mathcal{G}$ can be employed. The error term $\boldsymbol{\eta}_k$ models the prediction error in the estimation of the modal eigenvalues, assumed to follow a zero-mean joint Gaussian distribution with variance proportional to the measured eigenvalues. With this assumption, the SHM likelihood function is 
	\begin{linenomath}\begin{equation}
			f_{\boldsymbol{Z}_{SHM,k}|X_k}(\boldsymbol{Z}_{SHM,k}|X_k) = \prod_{m=1}^{N_m}N\left(\widetilde{\lambda}_{k_m}- \lambda_{k_m}^\mathcal{G}\left(X_k, E = \theta^{''}(\widetilde{T}_k)E_0\right); 0, c_{\lambda m}^2 \widetilde{\lambda}_{k_m}^2\right)
			\label{SHM_likelihood_function}
	\end{equation}\end{linenomath}	
	Visual inspections might be required for complementing the monitoring at certain time instances over the structural life-cycle. At a time instance $t_k$, when a visual inspection is performed, the complementary inspection measurement equation is
	\begin{linenomath}\begin{equation}
			Z_{insp,k} = X_k + \epsilon_k
			\label{process_equation_2}
	\end{equation}\end{linenomath}
	where $Z_{insp,k}$ is the uncertain outcome of a visual inspection and $X_k$ is the unknown true deterioration state at time $t_k$. $\epsilon_k$ models the uncertainty of the visual inspection outcome, and is assumed to follow a zero-mean Gaussian distribution with an assigned coefficient of variation ($\text{cv}_{insp}$) reflecting the quality of the inspection (in the numerical investigations of Section \ref{sec:Num_inv}, $\text{cv}_{insp}$=0.15). The resulting inspection likelihood function is
	\begin{linenomath}\begin{equation}
			f_{Z_{insp,k}|X_k}(Z_{insp,k}|X_k) = N\left(Z_{insp,k}; X_k, \text{cv}_{insp}\right)
			\label{insp_likelihood_function}
	\end{equation}\end{linenomath}	
	
	A Bayesian filter \cite{Doucet_2001, Sarkka_2013, tatsis_2022} can be implemented to solve the sequential Bayesian joint state-parameter estimation problem. The Bayesian filter implemented within the presented VoSHM framework is provided in the algorithmic summary of Section \ref{sec:Summary}. More specifically, we implement an on-line particle filter \cite{Kantas_2015}, which performs Gaussian mixture (GM)-based \cite{McLachlan_2007} resampling \cite{Kamariotis_2022_b} whenever the effective sample size drops below a user-defined threshold (step 8 in the algorithmic summary of Section \ref{sec:Summary}). The specific resampling scheme aims to counteract the issues of sample degeneracy and impoverishment that occur in on-line joint state-parameter estimation settings \cite{Sarkka_2013}. By running the filter, one obtains for each time step $t_k$ the one-step ahead predictive posterior distribution $\pi_{pos}(\widetilde{\boldsymbol{X}}_k\mid \boldsymbol{Z}_{1:k-1})$ and the filtered posterior distribution $\pi_{pos}(\widetilde{\boldsymbol{X}}_k\mid \boldsymbol{Z}_{1:k})$. It is not the primary focus of this paper to provide a thorough elaboration on the use of Bayesian filtering; to this end, the interested reader is referred to \cite{tatsis_2022, Kamariotis_2022_b}, where code is offered on an extended set of Bayesian algorithms. More specifically, the particle filter with GM resampling (PFGM) presented in \cite{Kamariotis_2022_b} is the method that we employ in this work.
	
	\subsection{Structural reliability and its updating}
	In many instances, a failure event at time $t$ can be expressed in terms of a structural capacity $R(t)$ and a demand $S(t)$, which are both random variables. We assume that the structural capacity $R(t)$ can be separated from the demand $S(t)$, and that a deterministic function $R(X_k)$ that outputs the structural capacity for a given deterioration state $X_k$ can be determined. Modeling $R(X_k)$ as deterministic is based on the assumption that the main uncertainties stem from the deterioration and the load, and that further physical and model uncertainties related to the structural capacity are not incorporated. More details on the definition of this problem-dependent deterministic function are provided for a specific structure in Section \ref{sec:Num_inv}, as well as in previous work of the authors \cite{Kamariotis_2022}. Such a simplified modeling choice is adopted here for the purpose of enabling a VoSHM analysis of practicable computational cost. The uncertain demand acting on the structure is modeled by the distribution of the maximum load $S_{max}$ in a one-year time interval. $F_{s_{max}}$ denotes the cumulative distribution function (CDF) of this distribution. 
	
	The time-variant reliability problem can be replaced by a series of time-invariant reliability problems \cite{Straub_2020}. $F_k^*$ is defined as the event of failure in interval $(t_{k-1}, t_k]$. For a given value of the deterioration state $X_k$, the structural capacity $R(X_k)$ is fixed and the conditional interval probability of failure is defined as:
	\begin{linenomath}\begin{equation}
			\text{Pr}(F_k^* \mid X_k)= \text{Pr}\left(S_{max}>R(X_k)\right) = 1 - \text{Pr}\left(S_{max}\leq R(X_k)\right) = 1 -F_{s_{max}}\big(R(X_k)\big)
	\end{equation}\end{linenomath}
	The unconditional accumulated probability of failure up to time $t_k$, $\text{Pr}(F_k) = \text{Pr}(F_1^*\cup F_2^*\cup...F_k^*)$, can be computed through the conditional interval probabilities $Pr(F_k^*|X_k)$,
	\begin{linenomath}\begin{equation}
			\text{Pr}(F_k|X_k) = 1-\prod_{m=1}^{k}\left[1 - \text{Pr}(F_m^*\mid X_m)\right]
			\label{conditional_accumulated}
	\end{equation}\end{linenomath}
	and by use of the total probability theorem:
	\begin{linenomath}\begin{equation}
			\text{Pr}(F_k) = \int_{\Omega_{X_k}}\text{Pr}(F_k\mid X_k) \pi(X_k)dX_k.
			\label{accumulated}
	\end{equation}\end{linenomath}
	If one replaces $\pi(X_k)$ with $\pi_{pos}(X_k\mid \boldsymbol{Z}_{1:k})$ in Equation \eqref{accumulated}, the updated estimate of the accumulated probability of failure is obtained:
	\begin{linenomath}\begin{equation}
			\text{Pr}(F_k|\boldsymbol{Z}_{1:k}) = \int_{\Omega_{X_k}}\text{Pr}(F_k\mid X_k) \pi_{pos}(X_k\mid \boldsymbol{Z}_{1:k})dX_k
			\label{accumulated_updated}
	\end{equation}\end{linenomath}
	The above integral can be solved with random sampling-based techniques. Here we employ a particle filtering scheme \cite{Sarkka_2013} to obtain weighted samples following $\pi_{pos}(X_k\mid \boldsymbol{Z}_{1:k})$.
	
	\section{Algorithmic summary of the heuristic-based expected total life-cycle cost calculation}
	\label{sec:Summary}
	
	In this section, we present a detailed algorithmic summary of the proposed methodology for the heuristic-based total expected life-cycle cost computation in the SHM preposterior analysis, i.e., the computation of $\boldsymbol{\text{E}}_{\boldsymbol{X}, \boldsymbol{Z}_{insp}, \boldsymbol{Z}_{SHM}}\left[C_{\text{tot}}\left(\boldsymbol{X}, \boldsymbol{Z}_{insp}, \boldsymbol{Z}_{SHM}, \boldsymbol{w}\right)\right]$. For the sake of readability, in this section $\boldsymbol{Z} = [\boldsymbol{Z}_{SHM}, \boldsymbol{Z}_{insp}]$.
	
	\begin{itemize}
		\item Fix the heuristic parameter vector $\boldsymbol{w} = [p_{th}^I, \Delta t_I, p_{th}^R]$ which defines the heuristic strategy $S_{\boldsymbol{w}}$.
		\item Draw $n_{\text{MCS}}$ samples of the parameter vector $[Q^{(i)}, H^{(i)}, U^{(i)}, Y^{(i)}, \tau^{(i)}]$ defining multiple potential underlying ``true" realizations of the environmental variability model. For each realization, create eigenvalue (modal) ``measurements" sampled at different temperatures in the initial undamaged structural state, and perform an offline Bayesian estimation, e.g., using the iTMCMC algorithm \cite{Betz_2016}, to learn the posterior distribution of the environmental model parameters and subsequently obtain $\theta^{''(i)}$.
		\item Draw $n_{\text{MCS}}$ samples from the prior model of the deterioration process, which define multiple potential underlying ``true" realizations of the deterioration process $\boldsymbol{\widetilde{X}}^{(i)}=[X_k^{(i)}, A^{(i)}, B^{(i)}]^\intercal, \quad i=1,..,n_{MCS}$.  
		\item For each $\theta^{''(i)}$ and $\boldsymbol{\widetilde{X}}^{(i)}$ realization, do the following:
		\begin{itemize}
			\item Draw $n_p$ particles of the initial deterioration state $X_0^{(j)}$ and the time-invariant parameter vector $[A_0^{(j)}, B_0^{(j)}]$ and set particle weights $w_0^{(j)}=\frac{1}{n_p}, \quad j=1,..,n_p$.
			\item For each $k=1,\dots, T$ do the following:
			\begin{enumerate}
				\item Draw a new point $X_k^{(j)}$ for each point in the particle set $\{X_{k-1}^{(j)}, j = 1,..,n_p\}$ from the deterioration state process equation:
				
				\medskip
				\begin{center}
					$X_k^{(j)} = X_{k-1}^{(j)} + A_{k-1}^{(j)}B_{k-1}^{(j)}\left(\frac{t_{k-1}+t_k}{2}\right)^{B_{k-1}^{(j)}-1}\Delta t \cdot \text{exp}(\omega_k) + \Delta D_k^{(j)}$
				\end{center}
				
				\medskip
				and for the time-invariant parameters:
				
				\medskip
				\begin{center}
					$[A_k^{(j)}, B_k^{(j)}]^\intercal = [A_{k-1}^{(j)}, B_{k-1}^{(j)}]^\intercal$
				\end{center}
				
				\medskip
				\item Estimate the one-step-ahead prediction for the time-dependent accumulated failure probability and the hazard function $h$, which expresses the failure rate of the structure conditional on survival up to the previous time instance:
				
				\medskip
				\begin{center}
					$Pr\left(F_k \mid \boldsymbol{Z}^{(i)}_{1:k-1}\right)\approx \sum_{j=1}^{n_p} w_{k-1}^{(j)} Pr(F_k \mid X_k^{(j)})$
				\end{center}
				
				\bigskip
				\begin{center}
					$h_k \left( \boldsymbol{Z}^{(i)}_{1:k-1}\right) \approx \frac{Pr\left(F_k \mid \boldsymbol{Z}^{(i)}_{1:k-1}\right) - {Pr\left(F_{k-1} \mid \boldsymbol{Z}^{(i)}_{1:k-1}\right)}}{Pr\left(F_{k-1} \mid \boldsymbol{Z}^{(i)}_{1:k-1}\right)}$
				\end{center}
				
				\medskip
				\item If $h_k \left( \boldsymbol{Z}^{(i)}_{1:k-1}\right)\geq p_{th}^R$: a repair is prescribed at time $t_{k-1}$. Set $\{X^{(j)}_{k-1}=0, \quad j=1,..,n_p\}$ and go back to step 1, i.e., after repair the structure is assumed to return back to its original undamaged state, and starts deteriorating anew.
				
				\medskip
				\item If $h_k \left( \boldsymbol{Z}^{(i)}_{1:k-1}\right)\geq p_{th}^I$: an inspection needs to be performed at time $t_{k-1}$. The uncertain inspection outcome $Z^{(i)}_{insp,k-1}$ is sampled from the inspection likelihood function of Equation \eqref{insp_likelihood_function}.
				
				\medskip
				\item If $(t_k = t_{cpp})$, i.e., if an extreme event has occurred, an inspection needs to be performed at time $t_{k}$. The uncertain inspection outcome $Z^{(i)}_{insp,k}$ is sampled from the inspection likelihood function of Equation \eqref{insp_likelihood_function}.
				
				\medskip
				\item Sample the measurement $\boldsymbol{Z}^{(i)}_{SHM,k}$ from the SHM system model (see Figure \ref{fig:SHM_sampling}). 
				
				\medskip
				\item Perform the filtering step. Update the weights
				
				\bigskip
				\begin{center}
					$w_k^{(j)} \propto f_{\boldsymbol{Z}_{SHM,k}|X_k}(\boldsymbol{Z}^{(i)}_{SHM,k}|X_k^{(i)}) \cdot 
					f_{\boldsymbol{Z}_{insp,k}|X_k}(\boldsymbol{Z}^{(i)}_{insp,k}|X_k^{(i)}) \cdot w_{k-1}^{(j)}$
				\end{center}	
				
				\bigskip
				and normalize these to sum to unity. Filter the accumulated probability of failure
				
				\medskip
				\begin{center}
					$Pr\left[F_k \mid \boldsymbol{Z}^{(i)}_{1:k}\right]\approx \sum_{j=1}^{n_p} w_{k}^{(j)} Pr[F_k \mid X_k^{(j)}]$
				\end{center}
				
				\medskip
				\item $N_{eff,k}= \frac{1}{\sum^{n_p}_{j=1}(w_k^{(j)})^2}\leq c\cdot n_p$ with $c \in [0,1]$ indicates sample degeneracy. To resolve that, fit a GM proposal distribution according to $\{\boldsymbol{X}_k^{(j)},w_k^{(j)}\}$ \cite{McLachlan_2007}, which approximates the current posterior, and sample $n_p$ new particles from this GM proposal distribution. Reset particle weights to $w_k^{(j)}=\frac{1}{n_p}$ \cite{Kamariotis_2022_b}.
			\end{enumerate}
			
			\item The lifetime inspection, repair and risk of failure costs corresponding to  $\theta^{''(i)}$ and $\boldsymbol{\widetilde{X}}^{(i)}$ are:
			
			\medskip 
			\begin{center}
				$C_{\text{I}}^{(i)} = \sum_{m=1}^{n_{\text{insp}}}\gamma\left(t_{\text{insp}}^{(m)}\right)\widetilde{c_{\text{I}}}$
			\end{center}
			\begin{center}
				$C_{\text{R}}^{(i)} = \sum_{m=1}^{n_{\text{rep}}}\gamma\left(t_{\text{rep}}^{(m)}\right)\widetilde{c_{\text{I}}}$
			\end{center}
			
			\begin{center}
				$R_{\text{F}}^{(i)}= \sum_{k=1}^{T}\gamma(t_k)\widetilde{c_{\text{F}}} \left\{ \text{Pr}[F_k|X_k^{(i)}] - \text{Pr}[F_{k-1}|X_{k-1}^{(i)}]\right\}$
			\end{center}
		\end{itemize}
		
		\medskip
		\item Compute the expected value:
		\begin{center}
			$\boldsymbol{\text{E}}_{\boldsymbol{X}, \boldsymbol{Z}_{insp}, \boldsymbol{Z}_{SHM}}\left[C_{\text{tot}}\left(\boldsymbol{X}, \boldsymbol{Z}_{insp}, \boldsymbol{Z}_{SHM}, \boldsymbol{w}\right)\right] \approx \frac{1}{n_{\text{MCS}}}\sum^{n_{\text{MCS}}}_{i=1}\left(C_{\text{I}}^{(i)}+C_{\text{R}}^{(i)} + R_{\text{F}}^{(i)}\right)$
		\end{center}
	\end{itemize}
	
	The methodology presented above can further be employed ``as is" with only few modifications for the computation of $\boldsymbol{\text{E}}_{\boldsymbol{X}, \boldsymbol{Z}_{insp}}\left[C_{\text{tot}}\left(\boldsymbol{X}, \boldsymbol{Z}_{insp}, \boldsymbol{w}\right)\right]$, i.e., for the preposterior analysis in the case of only inspections.

	\section{Numerical investigations}
	\label{sec:Num_inv}
	
	\begin{figure}[ht!]
		\centerline{
			\includegraphics[width=\textwidth]{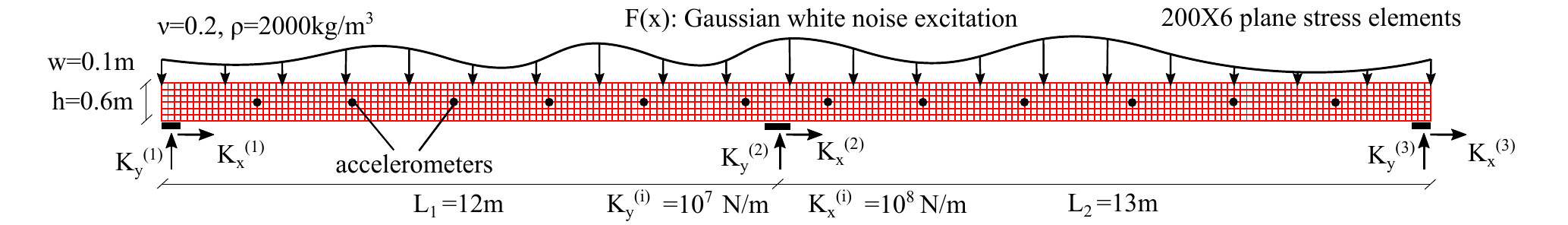}
		}
		\caption{Bridge system subject to environmental variability and damage due to deterioration at the middle pier.}
		\label{fig:benchmark}
	\end{figure}
	Figure \ref{fig:benchmark} shows the numerical benchmark model of a two-span bridge system \cite{Tatsis_2019}, which has already been employed by the authors for a VoI analysis in \cite{Kamariotis_2022, Kamariotis_2021}. The model is used as a simulator for creating dynamic response measurement samples from the bridge system, which is subjected to environmental variability and to gradual and shock deterioration at the middle elastic support, simulating the case of scour \cite{Prendergast_2013, Prendergast_2014, Garcia-Palencia_2015, Wang_2017, Sasidharan_2021}. Scour is one of the main causes of failure events on bridges \cite{Prendergast_2014, Sasidharan_2021}.
	
	Elastic boundaries in both directions are assumed for all three support points, in the form of translational springs with $K_x$ = 10\textsuperscript{8} N/m and $K_y$ = 10\textsuperscript{7} N/m. Damage is introduced as a reduction of the stiffness in the $y$-direction of the spring $K_y^{(2)}$. The evolution of damage over the bridge lifespan of $T = 50$ years is described by the damage model of Equation \eqref{damage}, where $K_{y,0}^{(2)}$ is the initial undamaged value, and $X(t)$ is the gradual and shock (e.g., due to flood occurrences) deterioration process, described by the model in Equation \eqref{deterioration_model}.
	\begin{linenomath}\begin{equation}
			K_y^{(2)}(t) = \frac{K_{y,0}^{(2)}}{1+X(t)}
			\label{damage}
	\end{equation}\end{linenomath}
	Modeling scour damage as a stiffness reduction at the support is not straightforward. When doing so, we ensure that the implemented damage properly reflects percentual changes of the modal properties equivalent to ones reported in literature for cases of scour \cite{Prendergast_2013, Prendergast_2014}. As a result, the modeled deterioration $X(t)$ can lead to large reductions of the stiffness $K_y^{(2)}(t)$.
	
	To model the environmental variability, a linear elastic material is assigned, with the Young's modulus assumed to vary with temperature, as described by Equation \eqref{E_eff}. The nominal value for the Young's modulus at a reference temperature of $20^{\circ}$C is $E_0=29.11$GPa, intended to represent reinforced concrete. The prior probabilistic models for the random variables entering models \eqref{E_eff}, \eqref{deterioration_model} are summarized in Table \ref{table:Parameters}.
	
	\begin{table}
		\caption{Prior distribution of environmental variability and deterioration model parameters}
		\footnotesize
		\centering
		\begin{tabular}{cccc}\hline
			Parameter&Distribution&Mean&cv\\\hline
			$Q$&Normal&-0.005&0.1\\
			$H$&Normal&1.115&0.025\\
			$U$&Normal&0.165&0.1\\
			$Y$&Normal&-1.00&0.25\\
			$\tau$ &Normal&3.00&0.20\\
			$A$&Lognormal&1.94$\cdot$10\textsuperscript{-4}&0.4\\
			$B$&Normal&2.0&0.10\\
			$\omega_k$&Normal&-0.005&0.10\\
			$D_i$&Lognormal&3.75&0.25\\
			$N(t)$&Poisson&0.04$\cdot$$t$&-\\\hline
		\end{tabular}
		\label{table:Parameters} 
	\end{table}
	
	\subsection{SHM probabilistic data sampling process}
	\label{subsec:SHM_sampling}
	
	\begin{figure}
		\centerline{
			\includegraphics[width=0.8\textwidth]{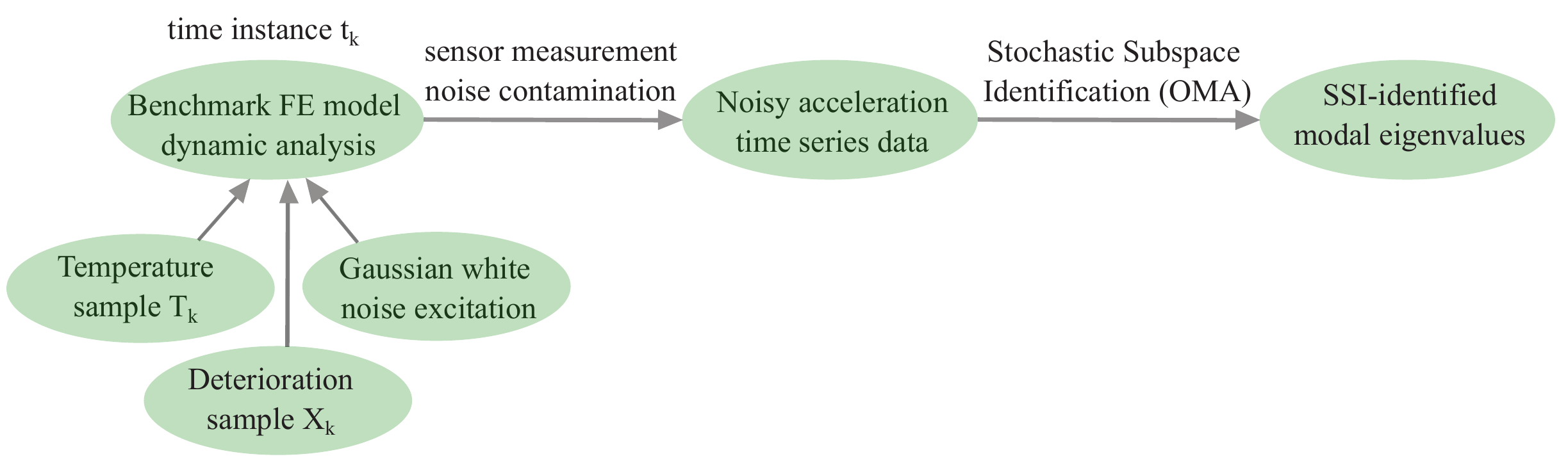}
		}
		\caption{SHM modal data sampling process}
		\label{fig:SHM_sampling}
	\end{figure}
	It is assumed that the two-span bridge system is continuously monitored from the beginning of its operation using a set of 12 sensors measuring vertical accelerations, whose locations correspond to predefined FE nodes (see Figure \ref{fig:benchmark}), as well as temperature sensors continuously providing ambient temperature measurements. A distributed Gaussian white noise excitation F(x) is used as the load acting on the bridge, to simulate the unknown ambient excitation. 
	
	As explained in Section \ref{sec:BA}, a VoSHM analysis requires a model of the SHM system, which can furnish probabilistic predictions of the life-cycle monitoring data. The SHM modal eigenvalue data sampling process is summarized in Figure \ref{fig:SHM_sampling}. At a time instance $t_k$, for a given input value of the deterioration state $X_k$ and for a given temperature $\widetilde{T}_k$, a dynamic time history analysis is run and the ``true" vertical acceleration signals $\ddot{x}$ at the sensor locations (FE nodes) are obtained. This noise-free acceleration time series data set is then contaminated with Gaussian white noise of 5\% root mean square noise-to-signal ratio, simulating a sensor measurement error. Subsequently, the noisy accelerations $\tilde{\ddot{x}}$ are fed into the SSI algorithm \cite{Peeters_1999}, which identifies between $m=3$ to $5$ lower eigenvalues.
	
	\subsection{Bayesian learning of the environmental variability model}
	
	This section demonstrates the Bayesian learning of the environmental variability model, as explained in subsection \ref{subsubsec:Bayesian_environmental}. The linear FE model $\mathcal{G}$ predicting the eigenvalues for the Bayesian updating process is the same FE model as the one used in the SHM modal data sampling process, which constitutes a so-called inverse crime \cite{Wirgin_2004}. However, this is a built-in feature of preposterior decision analysis. In a real SHM setting, the model is initially adjusted to reflect the true behavior as closely as possible, but existence of model uncertainty is inevitable, which will affect the accuracy of the VoSHM calculation. This, however, is a challenge pertinent to any engineering analysis.
	
\begin{figure}
		\begin{subfigure}{.33\textwidth}
			\centering
			\includegraphics[width=1.\linewidth]{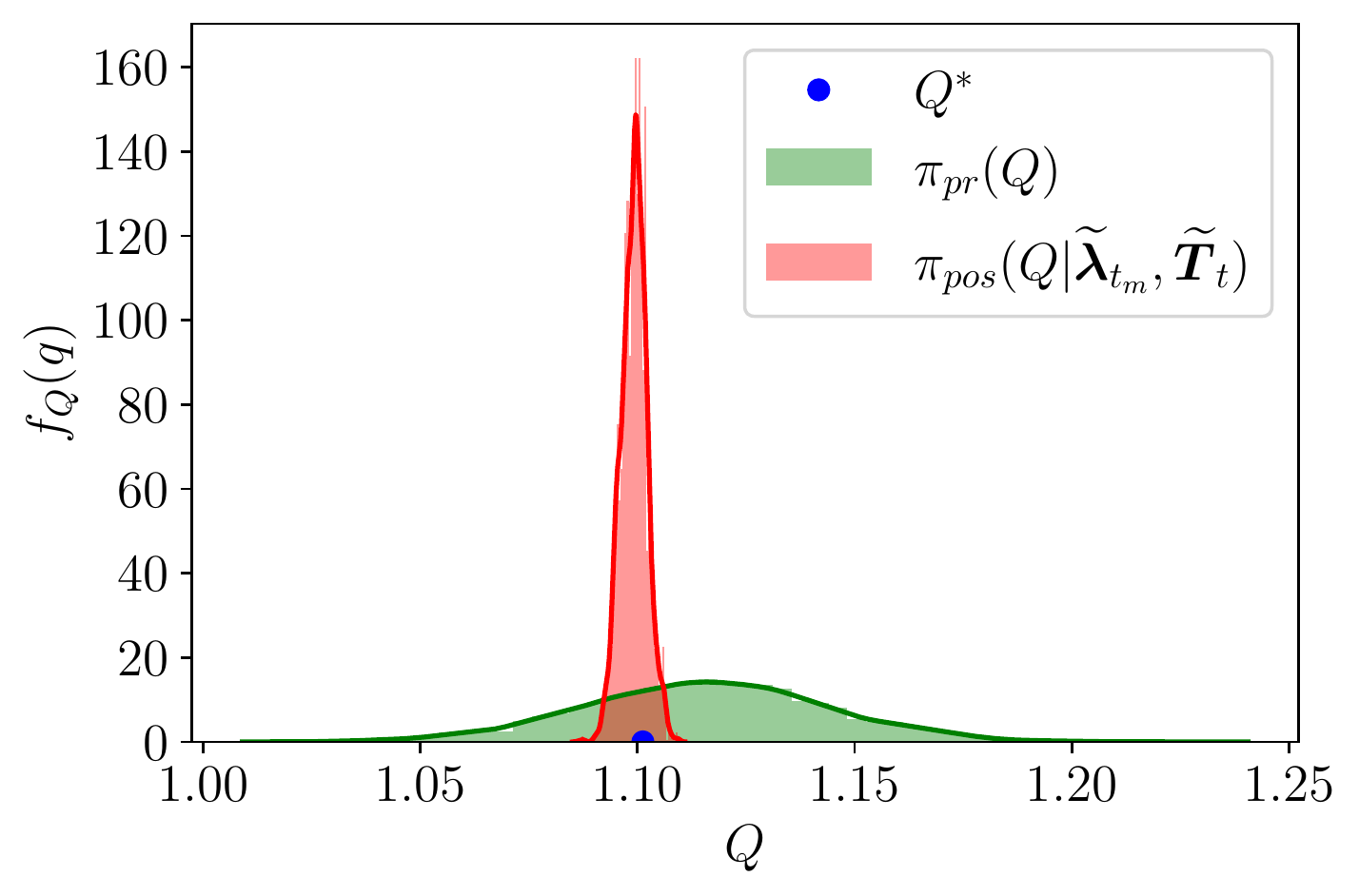}  
		\end{subfigure}
		\begin{subfigure}{.33\textwidth}
			\centering
			\includegraphics[width=1.\linewidth]{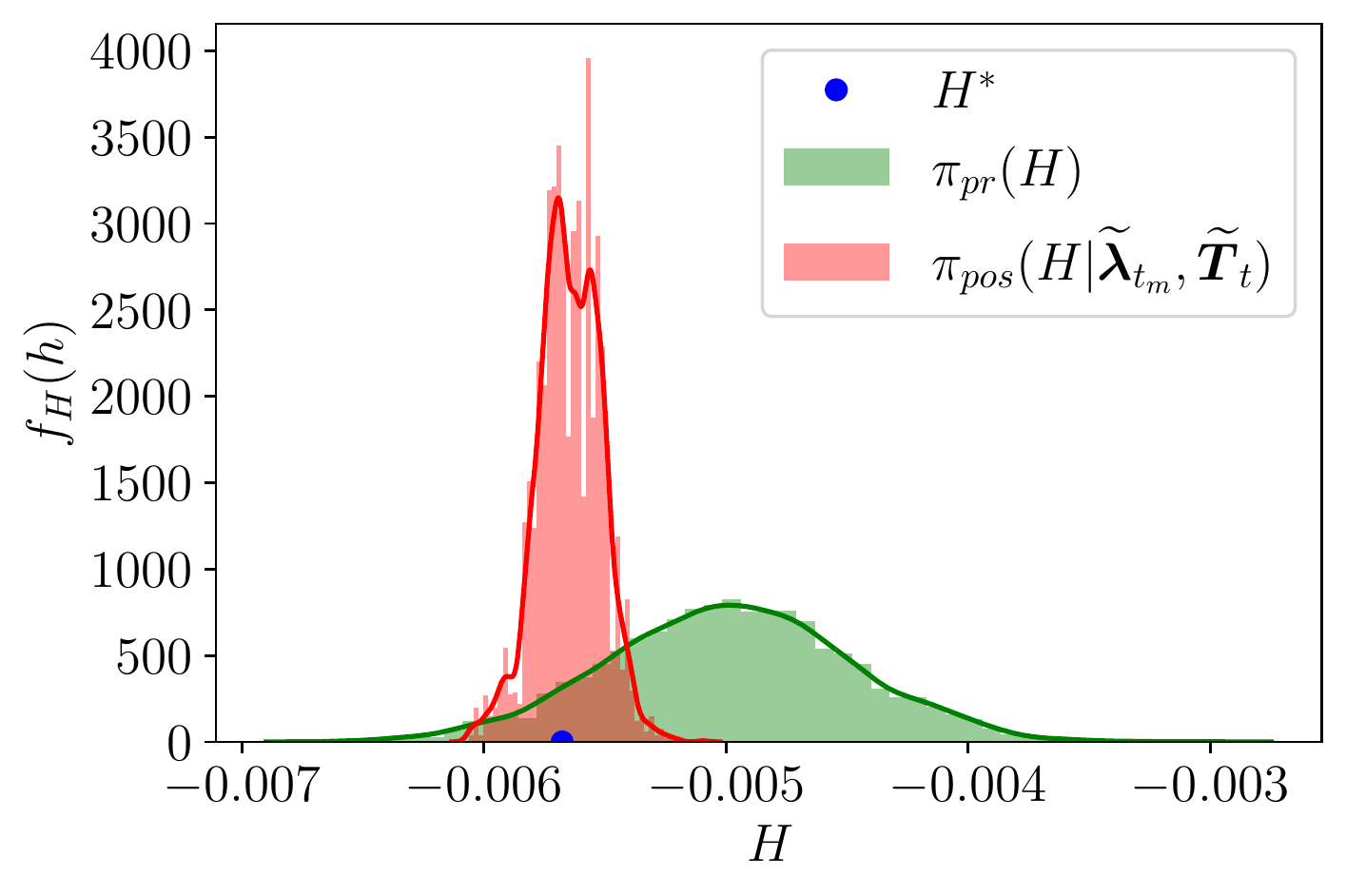}  
		\end{subfigure}
		\begin{subfigure}{.33\textwidth}
			\centering
			\includegraphics[width=1.\linewidth]{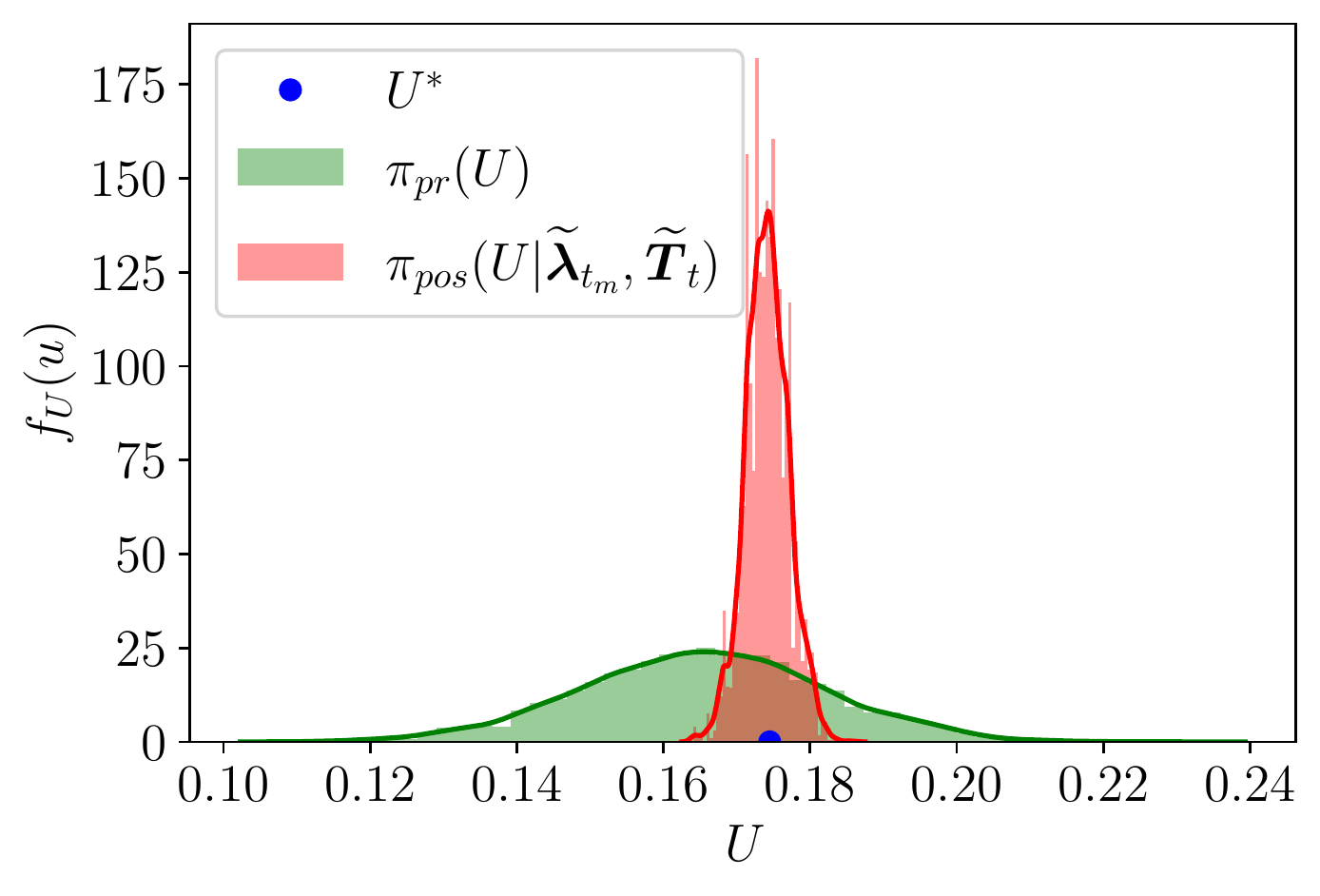}  
		\end{subfigure}
		\begin{subfigure}{.33\textwidth}
			\centering
			\includegraphics[width=1.\linewidth]{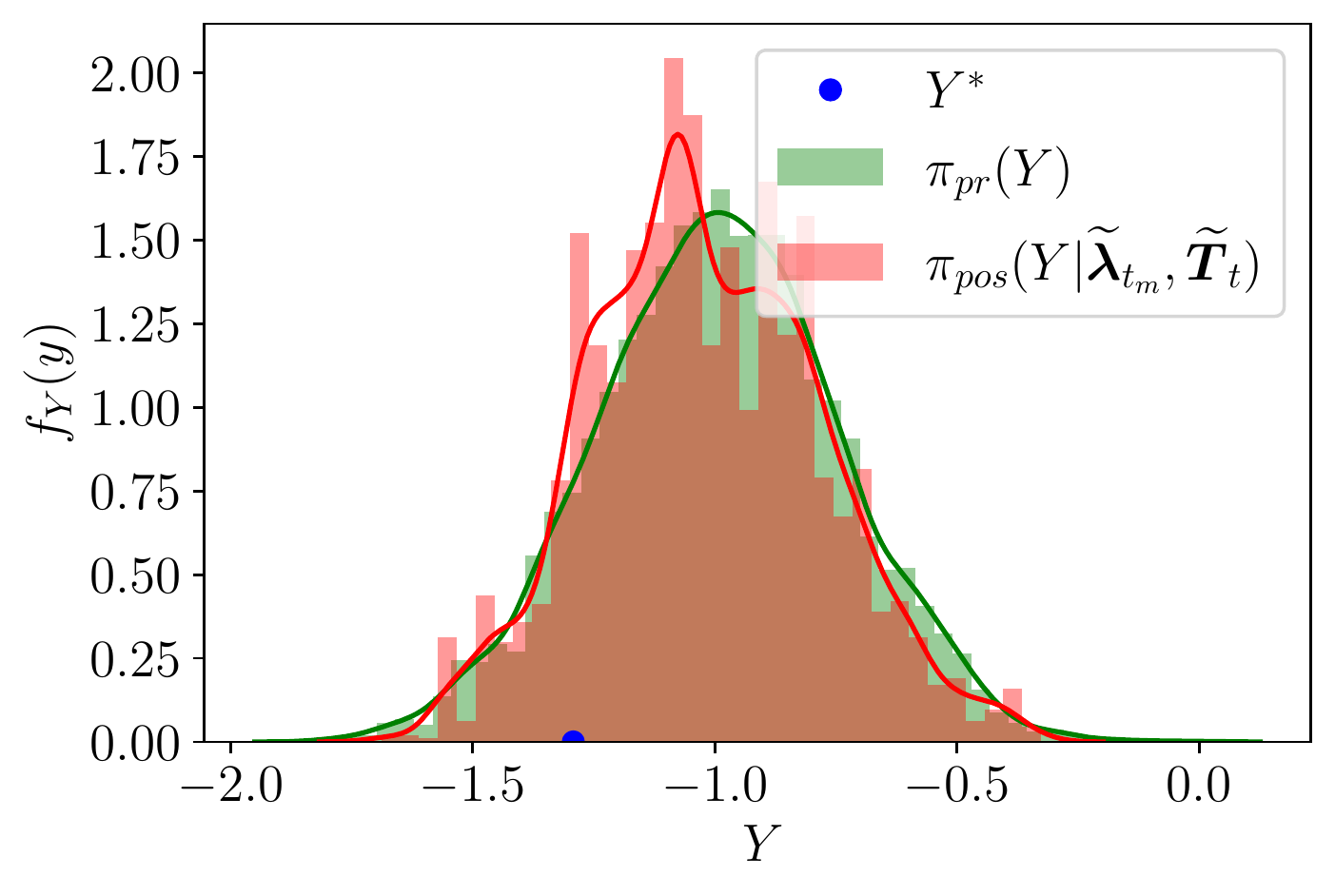}  
		\end{subfigure}
		\begin{subfigure}{.33\textwidth}
			\centering
			\includegraphics[width=1.\linewidth]{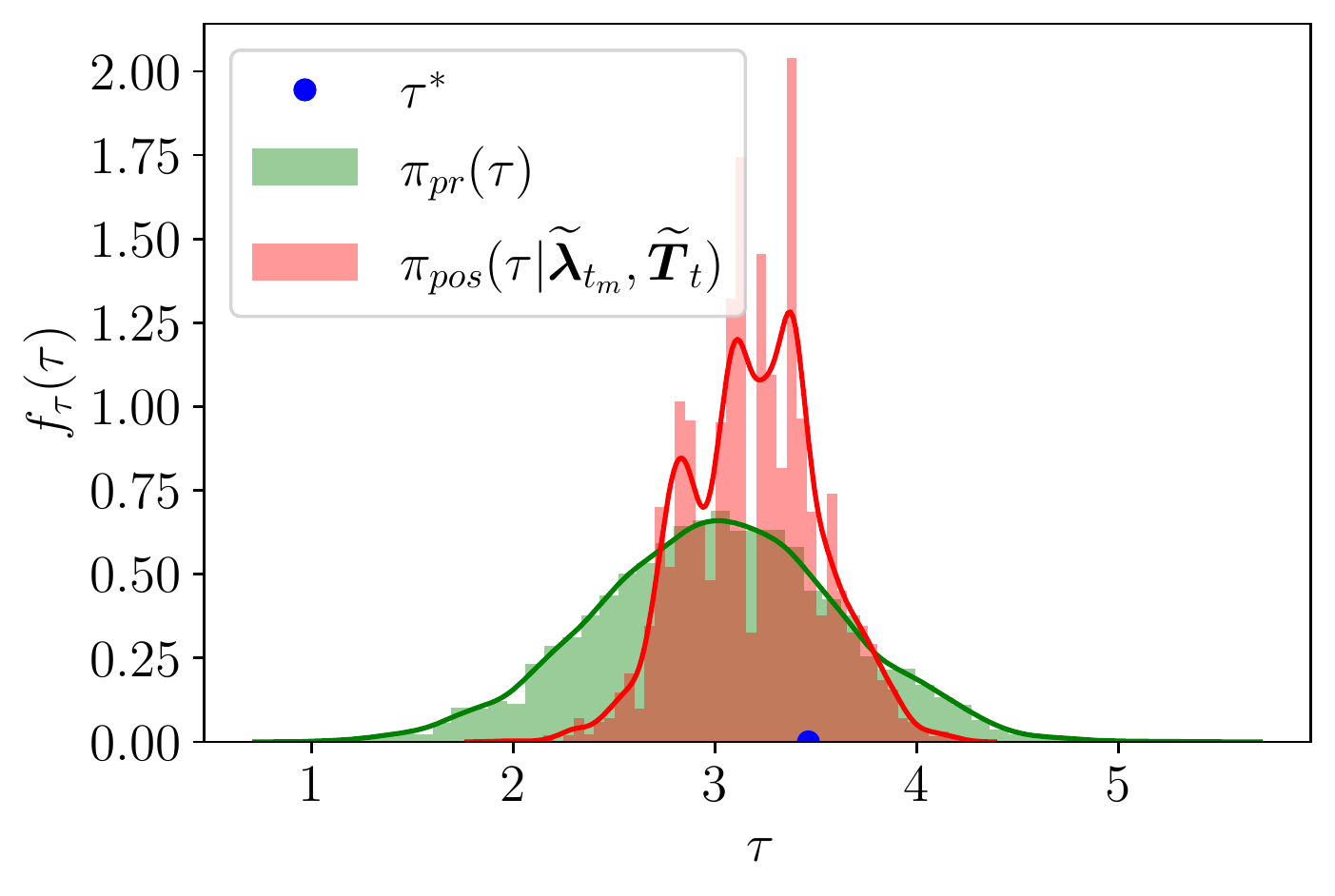}  
		\end{subfigure}
		\begin{subfigure}{.33\textwidth}
			\centering
			\includegraphics[width=1.\linewidth]{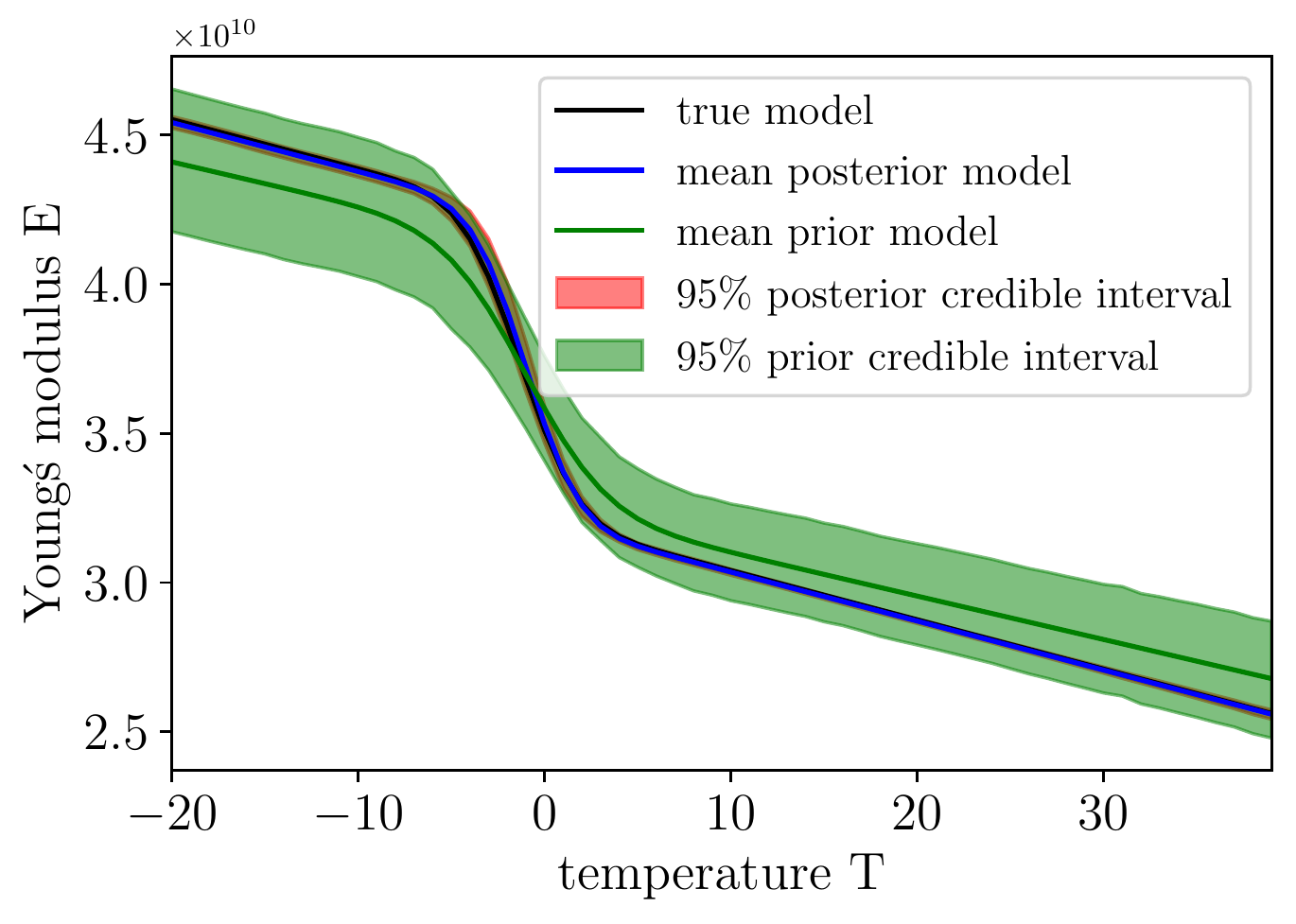}  
		\end{subfigure}
		\caption{Bayesian learning of the parameters of the environmental variability model.}
		\label{fig:updating_environmental}
	\end{figure}

	For illustrating the updating process, we consider one underlying ``true" realization of the environmental variability model, shown in black in the bottom right panel of Figure \ref{fig:updating_environmental}, with associated parameter values $[Q^*=-0.0057, H^*=1.101, U^*=0.174, Y^*=-1.292, \tau^*=3.464]$. We assume that $N_t$=50 eigenvalue sets, identified at different ambient temperature values via the SSI process, are available from the SHM in the initial operational period of the structure, when it is assumed that no damage is present. Via Bayesian analysis, the posterior distribution of the environmental variability model parameters is obtained. This is shown in Figure \ref{fig:updating_environmental}, where the first five panels plot the prior and posterior distribution of the five parameters. The last panel (bottom right) plots the prior mean model of Equation \eqref{E_eff} with its 95\% credible interval (CI), and the monitoring-informed estimate of the model $E(\widetilde{T}) = \theta^{''}(\widetilde{T}) \cdot E_0$ with its 95\% CI, obtained from posterior parameter samples generated from the iTMCMC algorithm \cite{Betz_2016}. One can observe that the underlying ``true" model is captured very well and the posterior 95\% CI is very narrow, reflecting small posterior uncertainty. For subsequent damage identification purposes, the learned blue model for the effective Young's modulus is employed.
	
	\subsection{VoSHM quantification}
	In Section \ref{subsec:heuristics} the heuristic-based VoSHM quantification via Equation \eqref{VoSHM_heuristic} is presented. For the numerical investigations in this section, the optimal heuristic thresholds $p_{th}^{I*}$ and $p_{th}^{R*}$ are computed via a preposterior analysis for the case of visual inspections only. The periodic inspections interval $\Delta t_I$ is set equal to 5 years, which is common practice for scour-specific inspections in many countries, and is not optimized, i.e., the heuristic parameters are $\boldsymbol{w}_1^*=[p_{th}^{I*}, p_{th}^{R*}, \Delta t_I=5]$. Obtaining $\boldsymbol{w}_2^*$ can become computationally unaffordable, therefore we use for $\boldsymbol{w}_2^*$ the same heuristic thresholds $p_{th}^{I*}$ and $p_{th}^{R*}$ that were optimized in the case of visual inspections only. This leads to a potential underestimation of the VoSHM, as these choices may not be fully optimal. In the case of continuous SHM, we assume that no fixed-interval periodic inspections will be performed. Naturally, this choice is based on the assumption of a high level of trust in the adequate functionality of the SHM system throughout the whole life-cycle. Eventually, $\boldsymbol{w}_2^*=[p_{th}^{I*}, p_{th}^{R*}, \Delta t_I=\infty]$.
	
	The following costs are assigned: $\widetilde{c}_f=5\cdot10^7$\euro, $\widetilde{c}_I=2\cdot10^4$\euro, $\widetilde{c}_R=6\cdot10^5$\euro. The scour inspection cost is assigned based on \cite{Briaud_2011}, and the scour repair cost based on \cite{Prendergast_2014, Anderson_2018}. The discount rate is taken as $r=2\%$. The lifetime of $T=50$ years is discretized into flexible decision-making intervals. ``Inspect?" \{yes, no\} and ``Repair?" \{yes,no\} decisions are made once per year, and potentially also at the specific time steps when an observed extreme event occurs.
	
	It should be noted that the model describing the outcome of a visual inspection that we assume in this work, described by Equation \eqref{insp_likelihood_function}, is a hypothetical, simplified model, employed for the sake of performing a VoSHM analysis. In reality, an inspection outcome would not come in the form of an estimated stiffness loss, and a more detailed analysis would be needed to express the inspection outcome in this form. Even so, this simplified inspection model can still incorporate the expected quality of the visual inspection by an appropriate choice of the assumed coefficient of variation $\text{cv}_{insp}$ in Equation \eqref{insp_likelihood_function}. The effect of this choice on the VoSHM result has been investigated by the authors in \cite{Kamariotis_2022_c}.
	
	For the structural reliability computation, the uncertain demand acting on the structure is modeled by the maximum load in each time interval with a Gumbel($a_n$=0.0509, $b_n$=0.297) distribution. The parameters of the Gumbel distribution are chosen such that the probability of failure in the initial undamaged state is equal to $10^{-6}$. In this work, we are interested in effects that are related to damage and are imprinted on monitored structural properties, in the form of residual stiffness reduction. We determine the deterministic function of the capacity for given deterioration state, $R(X_k)$, considering that for increasing stiffness reduction at the middle elastic support, the load bearing capacity of the bridge system decreases due to increase in the normal stresses at the middle of the right midspan. A one-dimensional grid of possible values of the deterioration $X_k$ is created, and each of those values is given as input for a static analysis with the FE model. For each implemented $X_k$ value, the loss of load bearing capacity of the structure relative to the initial undamaged state is evaluated, leading to the corresponding $R(X_k)$ value. The same modeling choice was employed by the authors in previous work (see Fig. 15 in \cite{Kamariotis_2022}).
	
\begin{figure}
		\centering
		\begin{subfigure}{.415\textwidth}
			\centering
			\includegraphics[width=1.\linewidth]{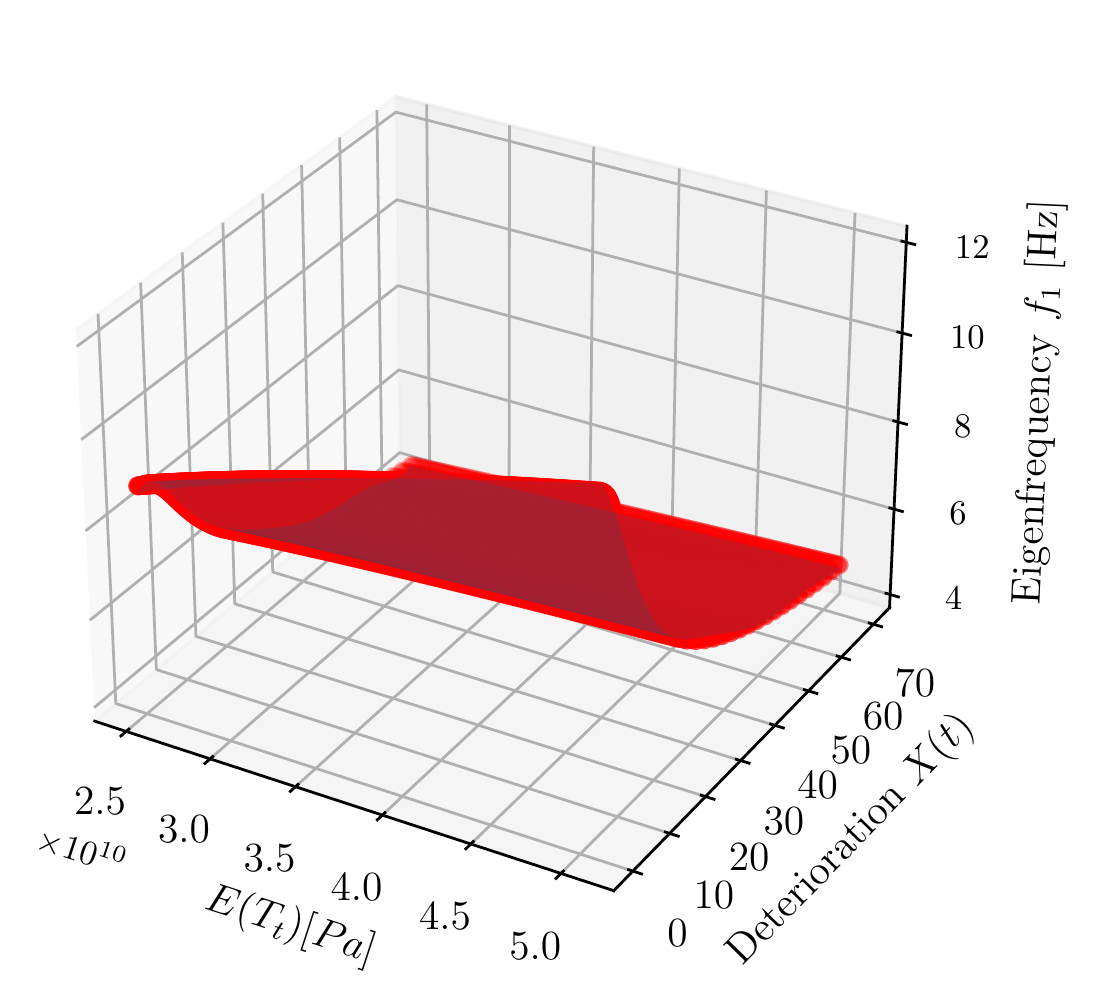}  
		\end{subfigure}
		\begin{subfigure}{.415\textwidth}
			\centering
			\includegraphics[width=1.\linewidth]{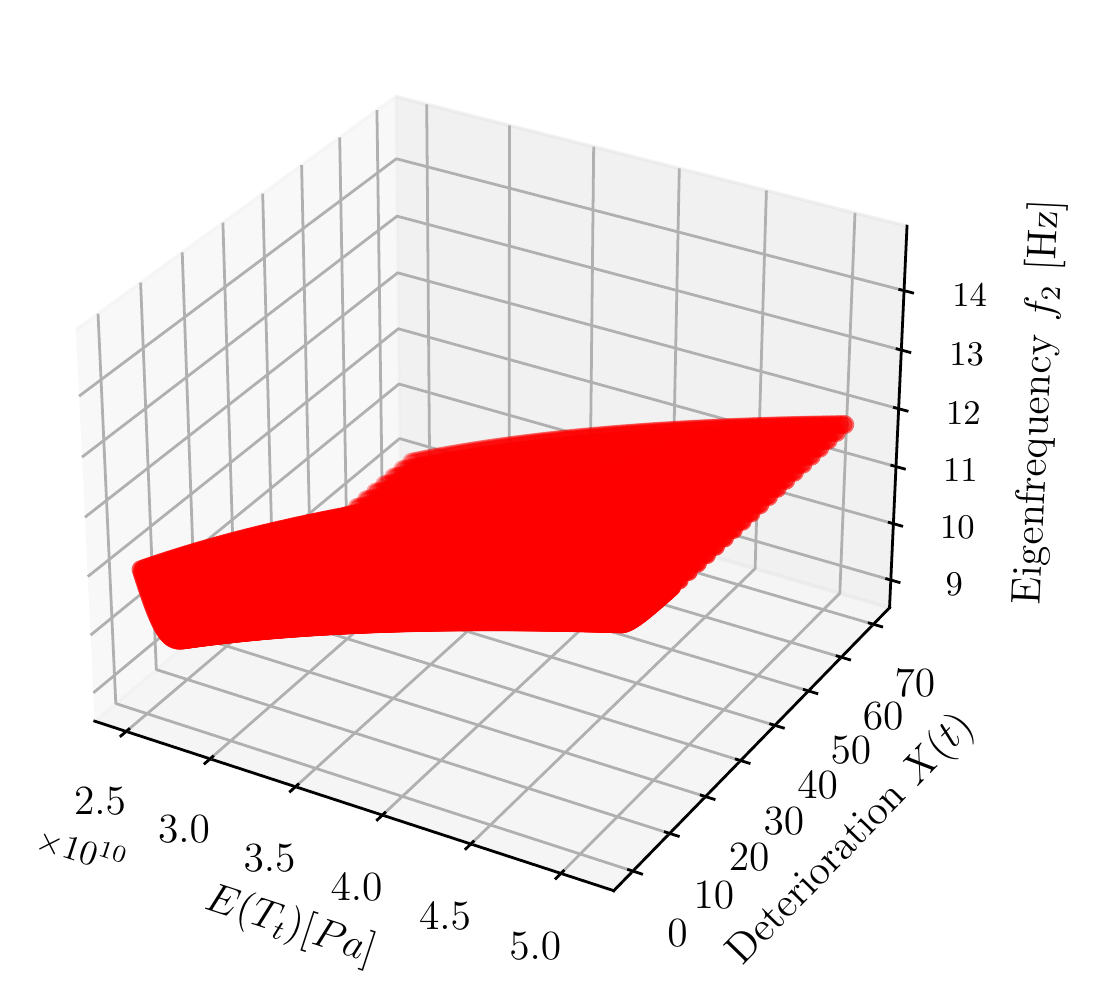} 
		\end{subfigure}
		\caption{Polynomial ridge regression surrogate model for eigenfrequencies $f_1$ and $f_2$.}
		\label{f:surrogate}
	\end{figure}

	Finally, for the sequential Bayesian estimation of the deterioration state and parameters with continuous vibration-based SHM data, a polynomial ridge regression \cite{Murphy_2012} surrogate model is used to replace each of the structural FE model $\mathcal{G}$-predicted eigenfrequencies entering Equation \eqref{process_equation}. We create a two-dimensional grid of possible values for the deterioration $X(t)$ and for the effective Young's modulus as a function of temperature $E(T_t)$. For each point in this two-dimensional grid, we execute a modal analysis using the structural FE model and store the output eigenfrequencies. Eventually, for each eigenfrequency, we fit a two-dimensional polynomial ridge regression response surface model, as exemplarily shown in Figure \ref{f:surrogate} for the first two eigenfrequencies, which we use as the surrogate model.
	
	\subsubsection{First case study: Gradual deterioration and shock deterioration due to an observed extreme event}
	\label{subsubsec:First}
	
	As a first case study, we assume a scenario with gradual and shock deterioration, where a CPP shock deterioration occurrence corresponds to an observed extreme event, e.g., a flood occurrence. In this case, both without and with continuous SHM, an additional visual inspection will take place right after the extreme event.
	
	Performing a heuristic-based expected total life-cycle cost minimization without SHM, the optimal heuristic parameter vector $\boldsymbol{w}_1^*=[p_{th}^{I*} = 5\cdot10^{-4}, p_{th}^{R*}=1\cdot10^{-3}, \Delta t_I=5]$ is found. For the solution of the joint expectation $\boldsymbol{\text{E}}_{\boldsymbol{X}, \boldsymbol{Z}_{insp}}$, as summarized in Section \ref{sec:Summary}, $n_{MCS}=1000$ samples were drawn, each defining one potential underlying ``true" realization of the deterioration process and the corresponding visual inspection data.
	
\begin{figure}[ht!]
		\centering
		\begin{subfigure}{.40\textwidth}
			\centering
			\includegraphics[width=1.\linewidth]{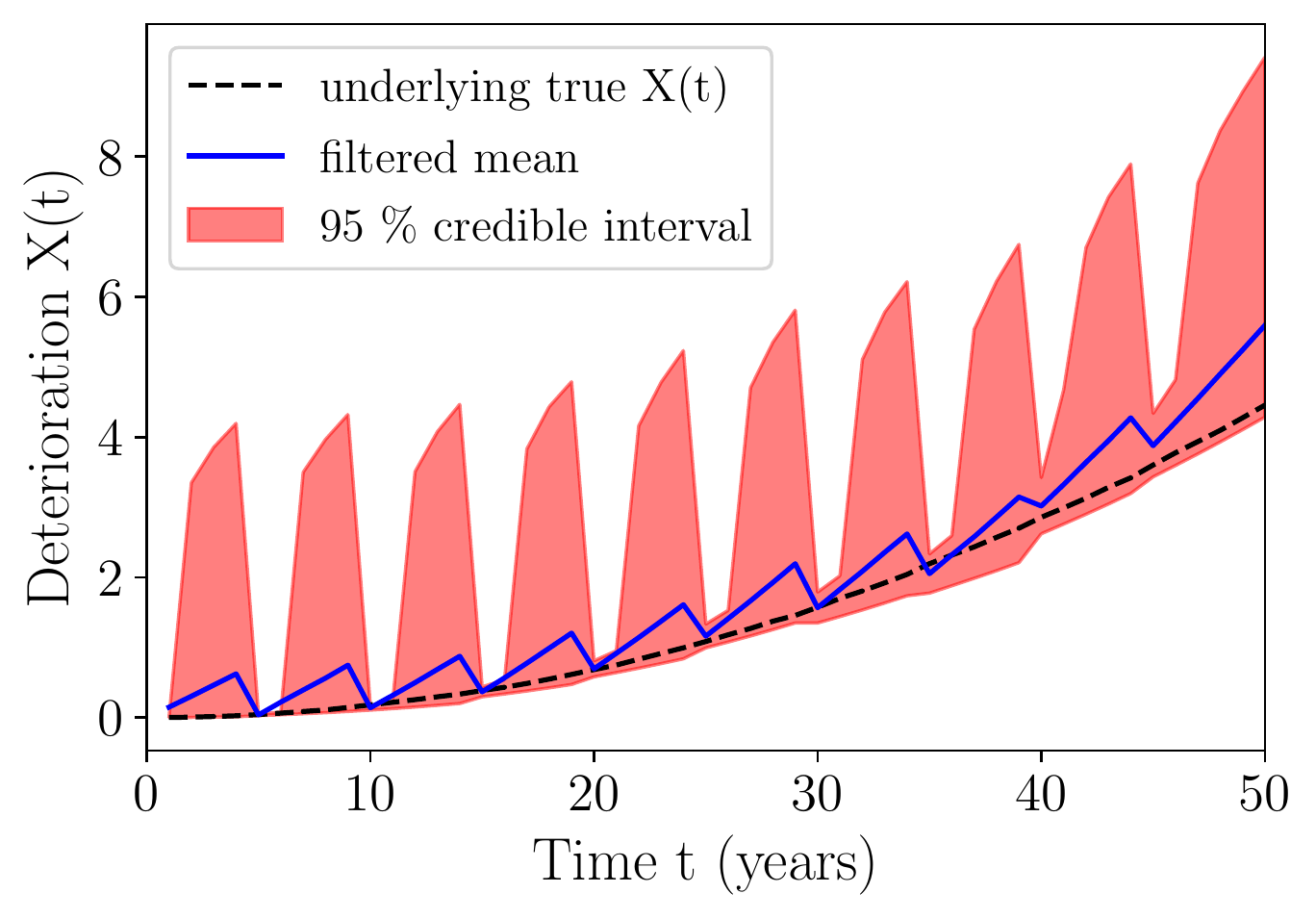}  
		\end{subfigure}
		\begin{subfigure}{.415\textwidth}
			\centering
			\includegraphics[width=1.\linewidth]{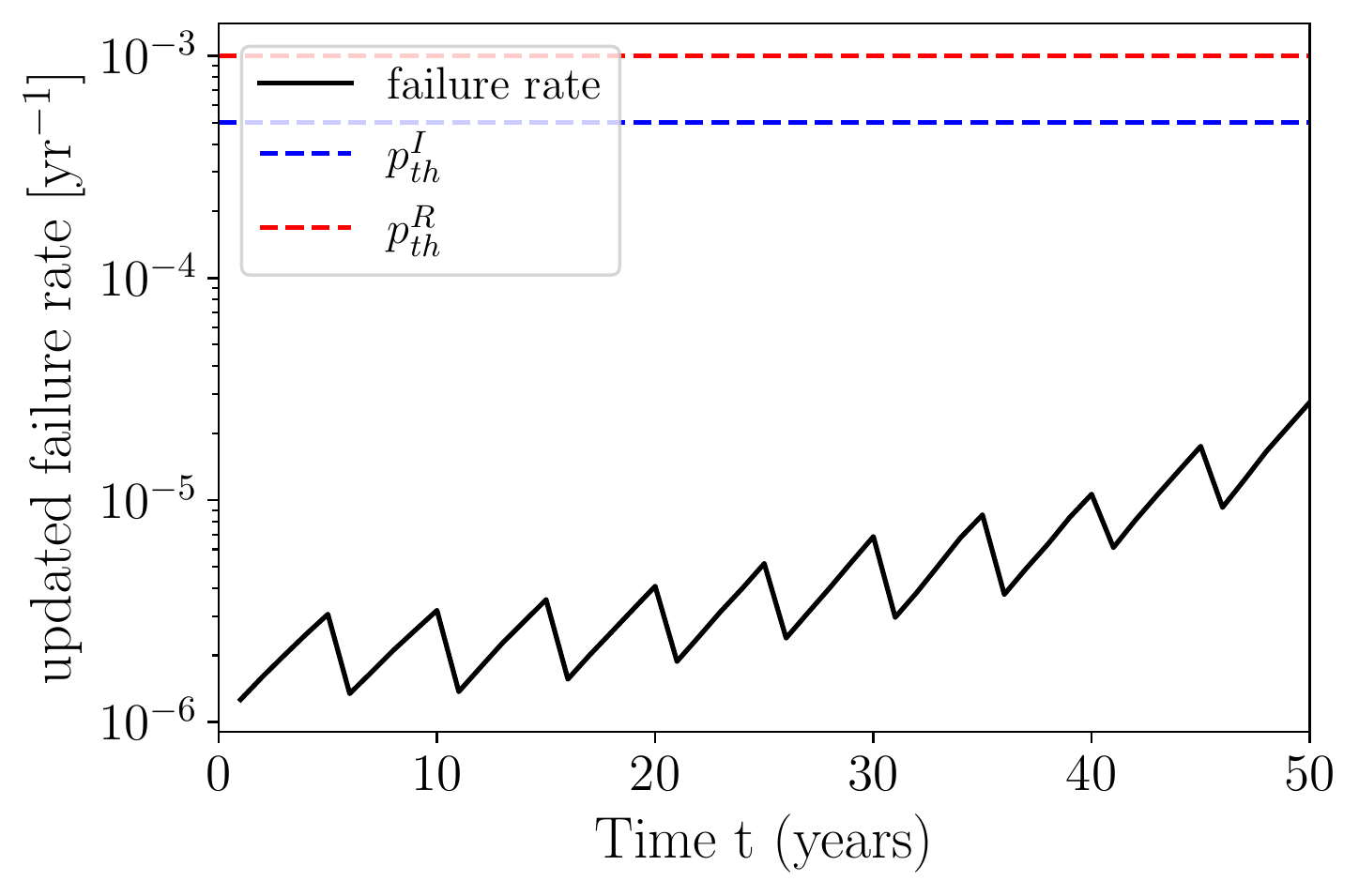} 
		\end{subfigure}
		\caption{Bayesian filtering of the deterioration state and the reliability using intermittent visual inspection data.}
		\label{f:1st_case_study_inspect}
	\end{figure}

\begin{figure}[ht!]
		\centering
		\begin{subfigure}{.40\textwidth}
			\centering
			\includegraphics[width=1.\linewidth]{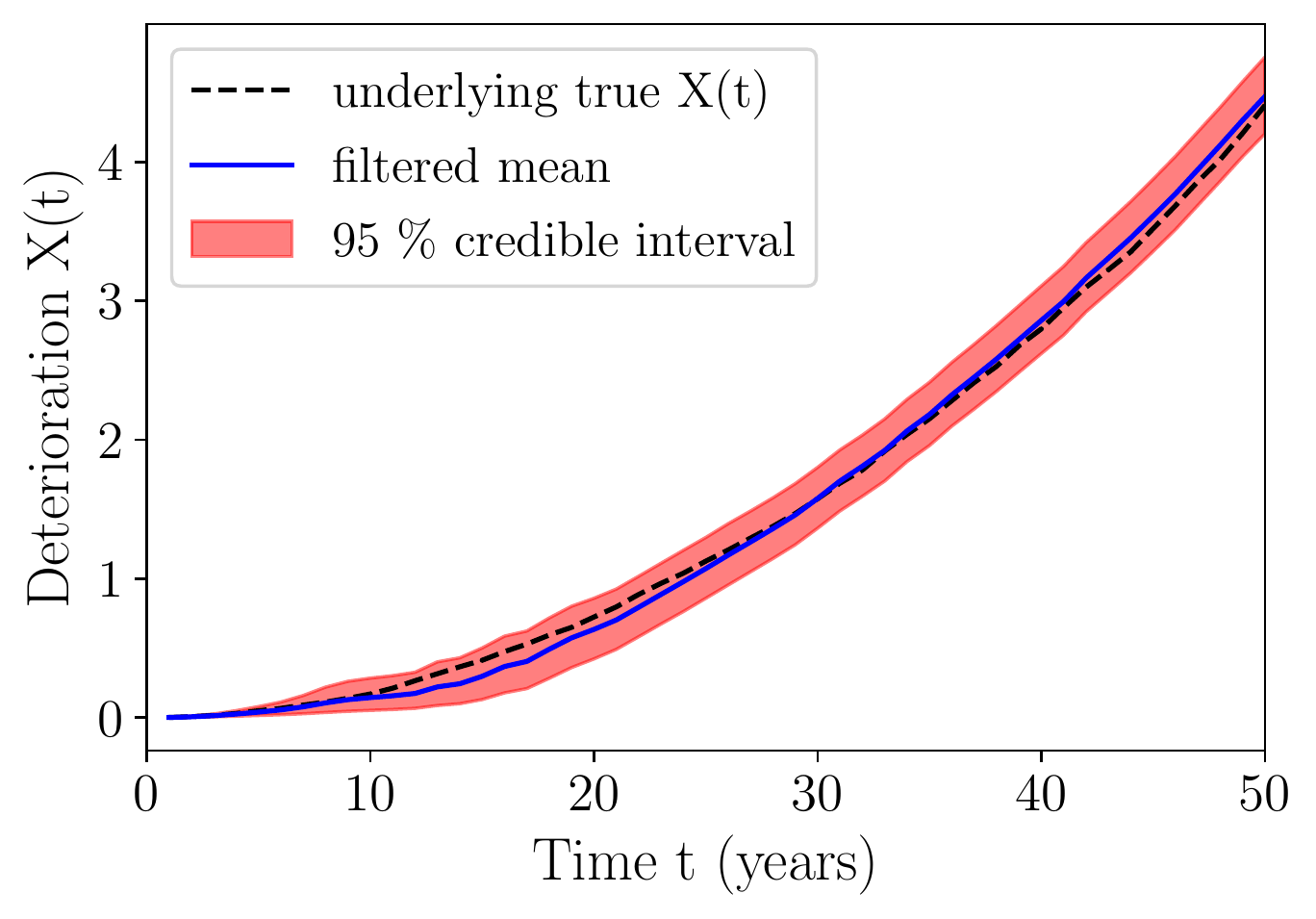}  
		\end{subfigure}
		\begin{subfigure}{.415\textwidth}
			\centering
			\includegraphics[width=1.\linewidth]{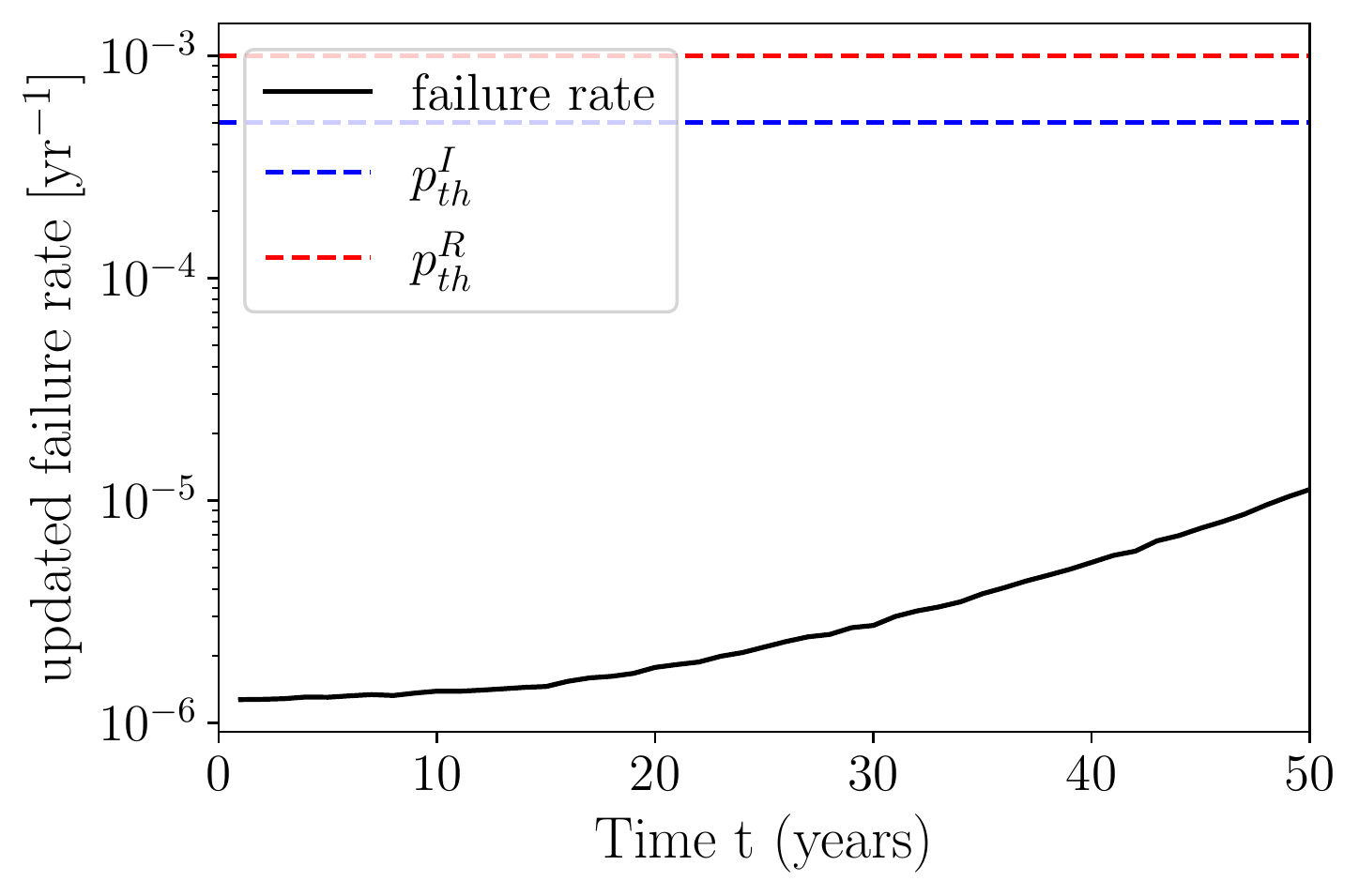} 
		\end{subfigure}
		\caption{Bayesian filtering of the deterioration state and the reliability using continuous SHM and inspection data.}
		\label{f:1st_case_study_SHM}
	\end{figure}

	We demonstrate how the sequential Bayesian estimation of the deterioration state and the corresponding sequential decision-making operates without and with continuous vibration-based SHM, by looking at one of these underlying ``true" deterioration process realizations. The dashed black line in the first panel of Figures \ref{f:1st_case_study_inspect}, \ref{f:1st_case_study_SHM} corresponds to this single underlying ``true" realization. For this sample, no extreme event occurs, and the deterioration process is driven by gradual deterioration only. The left panel of Figure \ref{f:1st_case_study_inspect} plots the filtered deterioration state estimate, and the right panel plots the filtered failure rate estimate, as obtained in view of intermittent visual inspection data. For $\boldsymbol{w}_1^*=[p_{th}^{I*} = 5\cdot10^{-4}, p_{th}^{R*}=1\cdot10^{-3}, \Delta t_I=5]$, a visual inspection is performed at $t_{insp} = [5, 10, 15, 20, 25, 30, 35, 40, 45]$ years, as dictated by $\Delta t_I=5$. Since $p_{th}^{I*}$ and $p_{th}^{R*}$ are not exceeded, no additional inspection or repair takes place. Figure \ref{f:1st_case_study_SHM} plots the corresponding estimates in the case when eigenvalue data from the investigated SHM system is continuously available. One can see that, in the presence of continuous SHM data, system state awareness at all times is accomplished, as opposed to Figure \ref{f:1st_case_study_inspect}, and the choice not to perform the periodic visual inspections leads to cost savings. More specifically, for this single realization $C_{\text{tot}}\left(\boldsymbol{x}, \boldsymbol{z}_{insp}, \boldsymbol{w}_1^*\right) - C_{\text{tot}}\left(\boldsymbol{x}, \boldsymbol{z}_{insp}, \boldsymbol{z}_{SHM}, \boldsymbol{w}_2^*\right)=1.13\cdot10^5$\euro. 
	
\begin{figure}[ht!]
		\centering
		\begin{subfigure}{.40\textwidth}
			\centering
			\includegraphics[width=1.\linewidth]{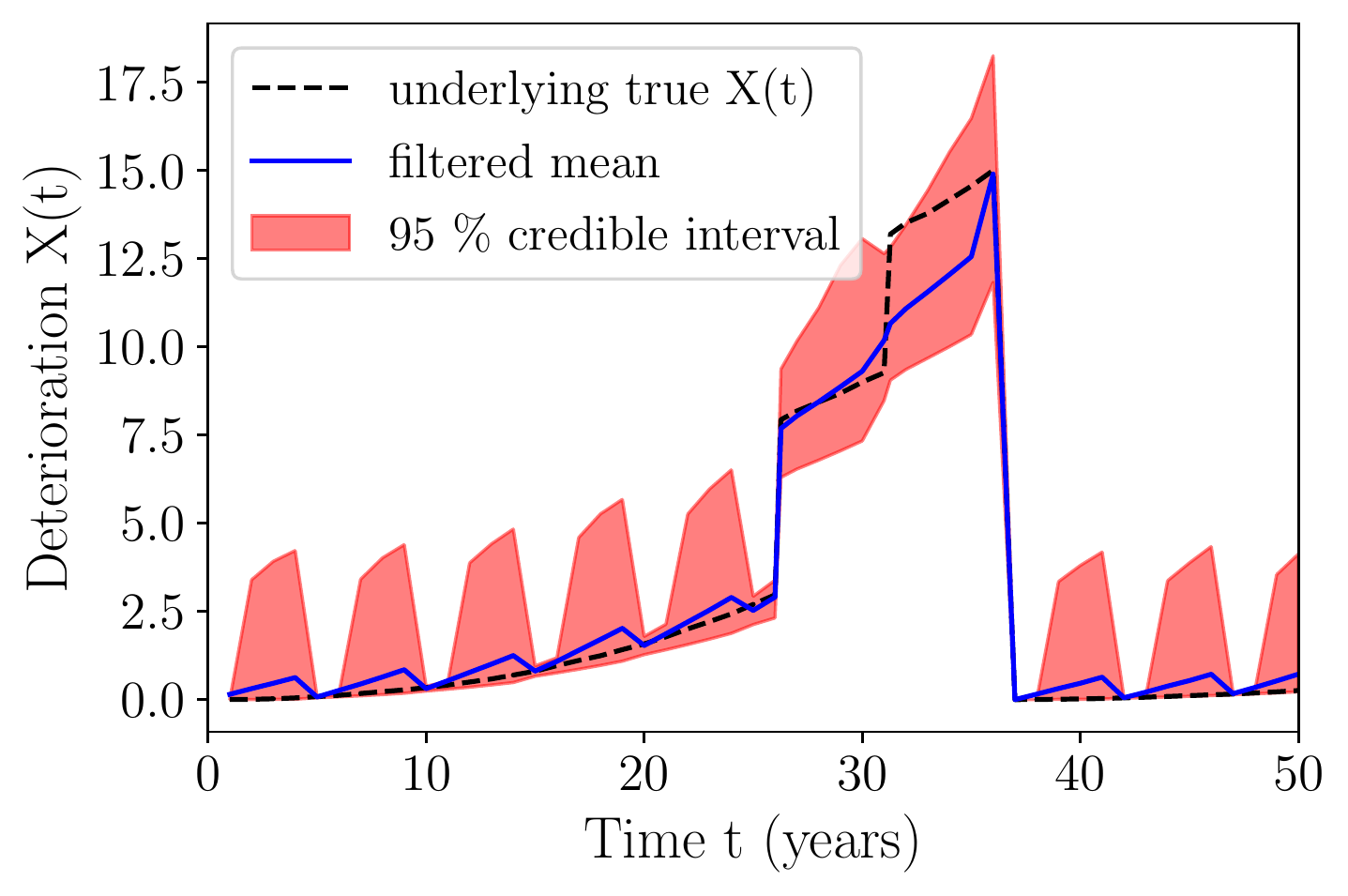}
		\end{subfigure}
		\begin{subfigure}{.415\textwidth}
			\centering
			\includegraphics[width=1.\linewidth]{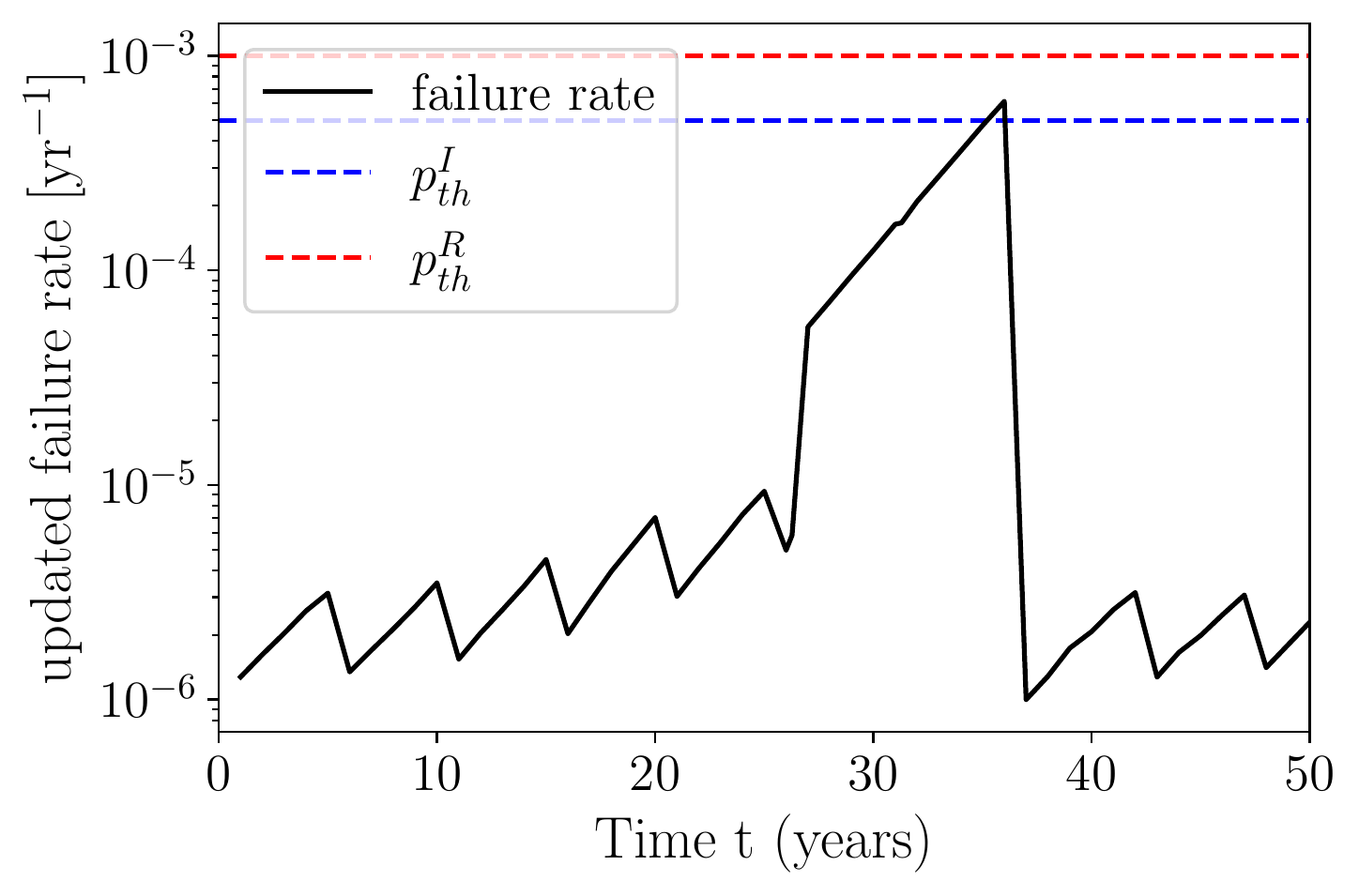}  
		\end{subfigure}
		\caption{Bayesian filtering of the deterioration state and the reliability using intermittent visual inspection data.}
		\label{f:1st_case_study_inspect_2}
	\end{figure}

\begin{figure}[ht!]
		\centering
		\begin{subfigure}{.40\textwidth}
			\centering
			\includegraphics[width=1.\linewidth]{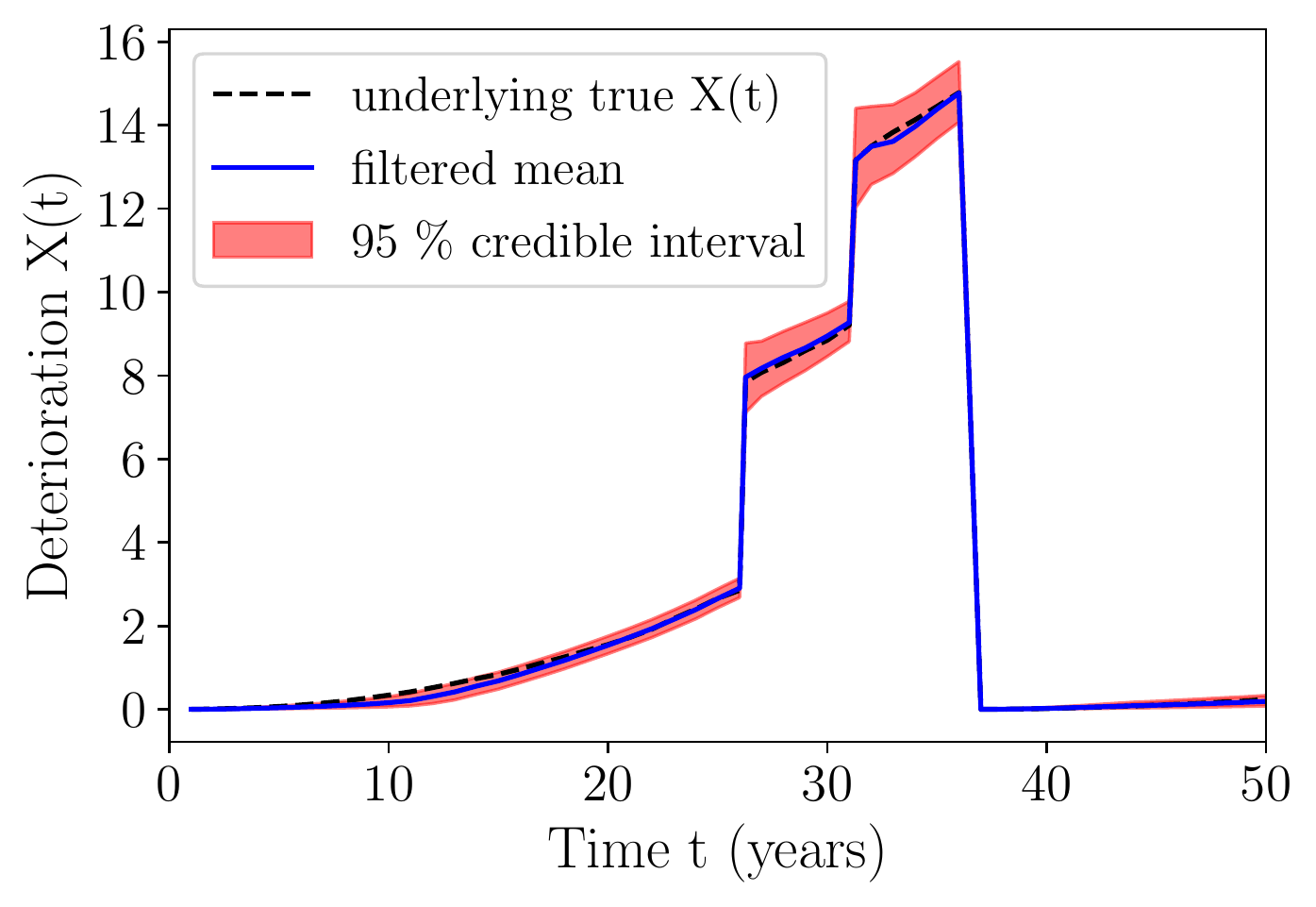}  
		\end{subfigure}
		\begin{subfigure}{.415\textwidth}
			\centering
			\includegraphics[width=1.\linewidth]{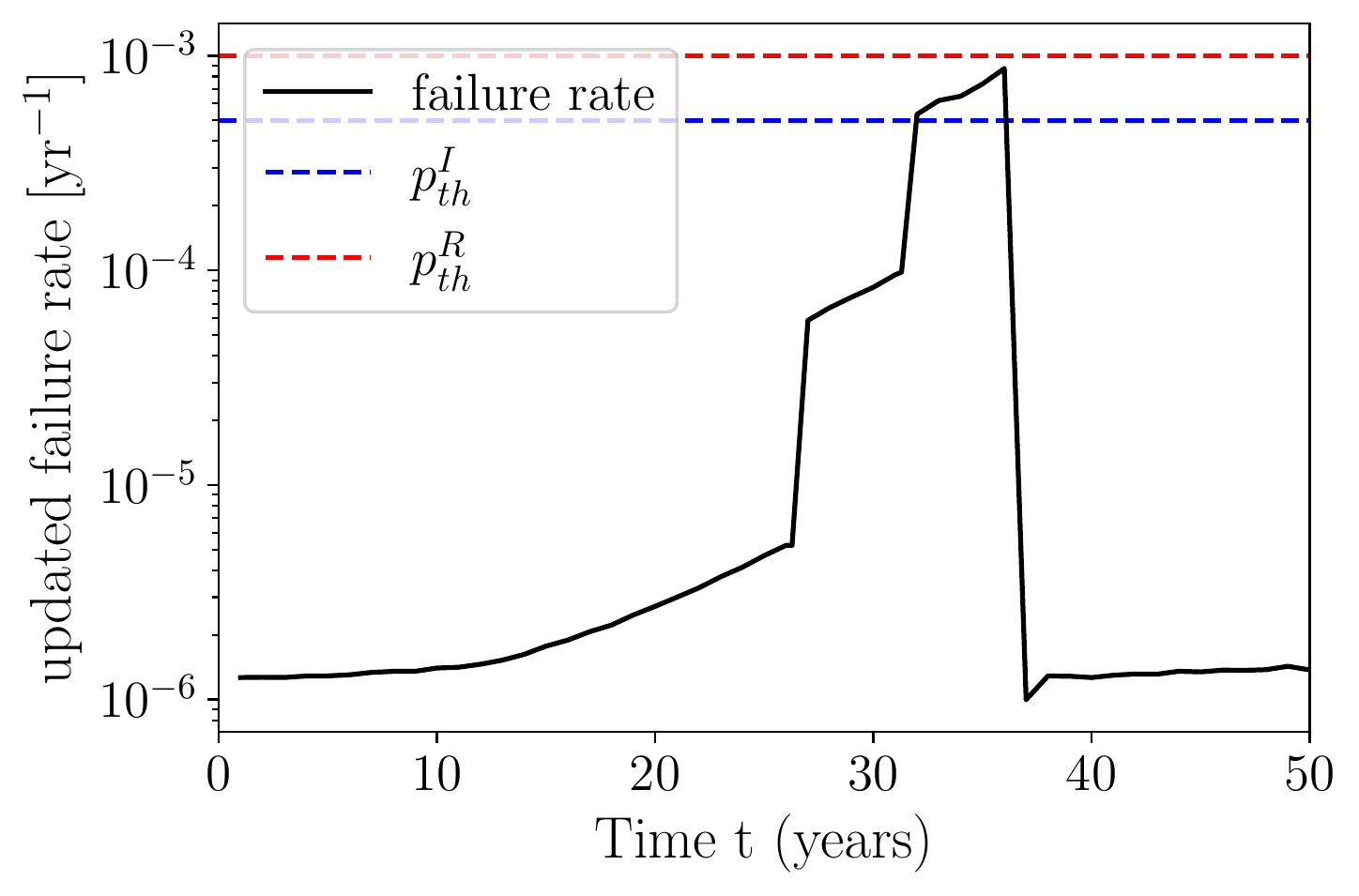}
		\end{subfigure}
		\begin{subfigure}{.415\textwidth}
			\centering
			\includegraphics[width=1.\linewidth]{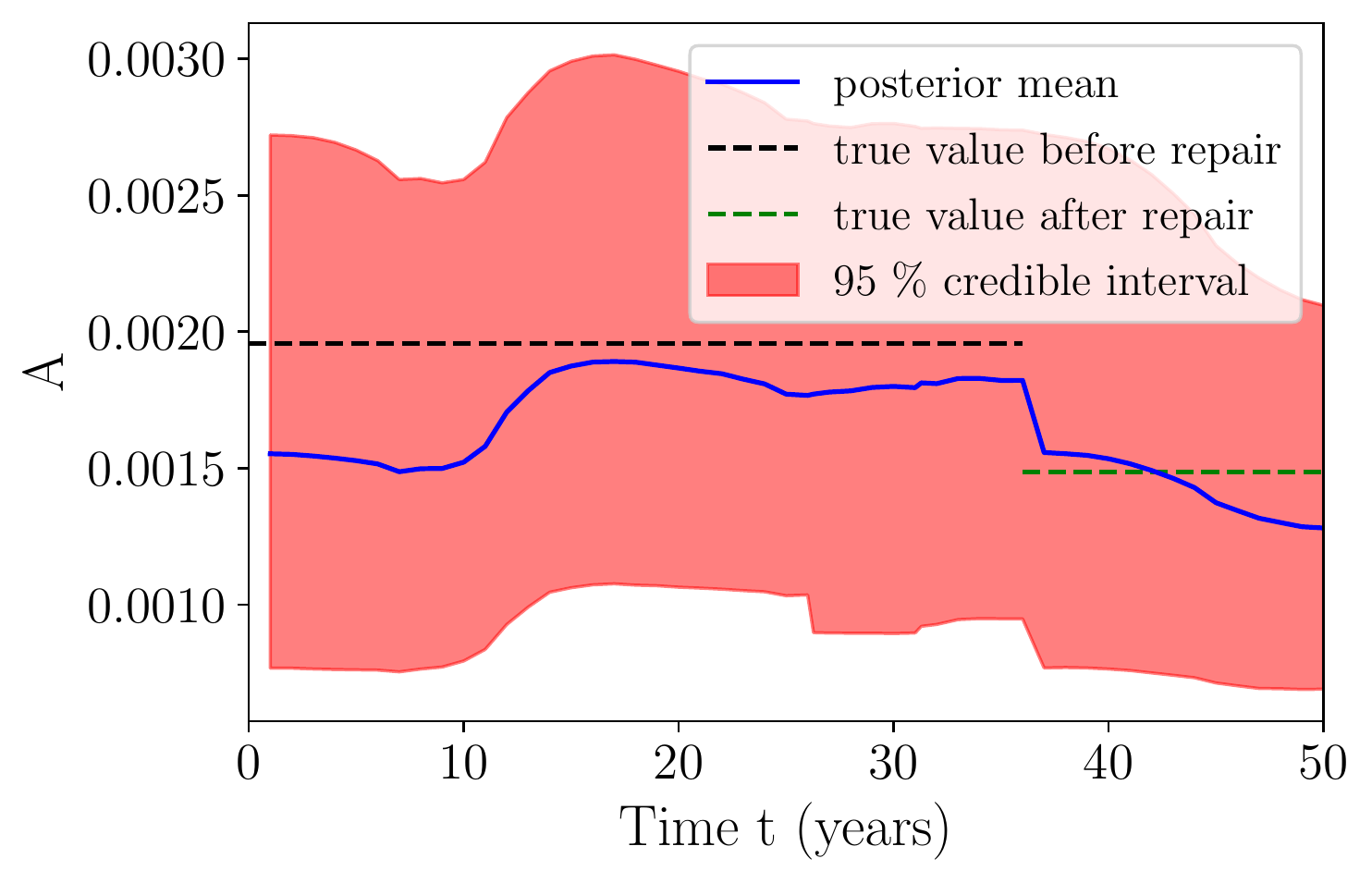}  
		\end{subfigure}
		\begin{subfigure}{.4\textwidth}
			\centering
			\includegraphics[width=1.\linewidth]{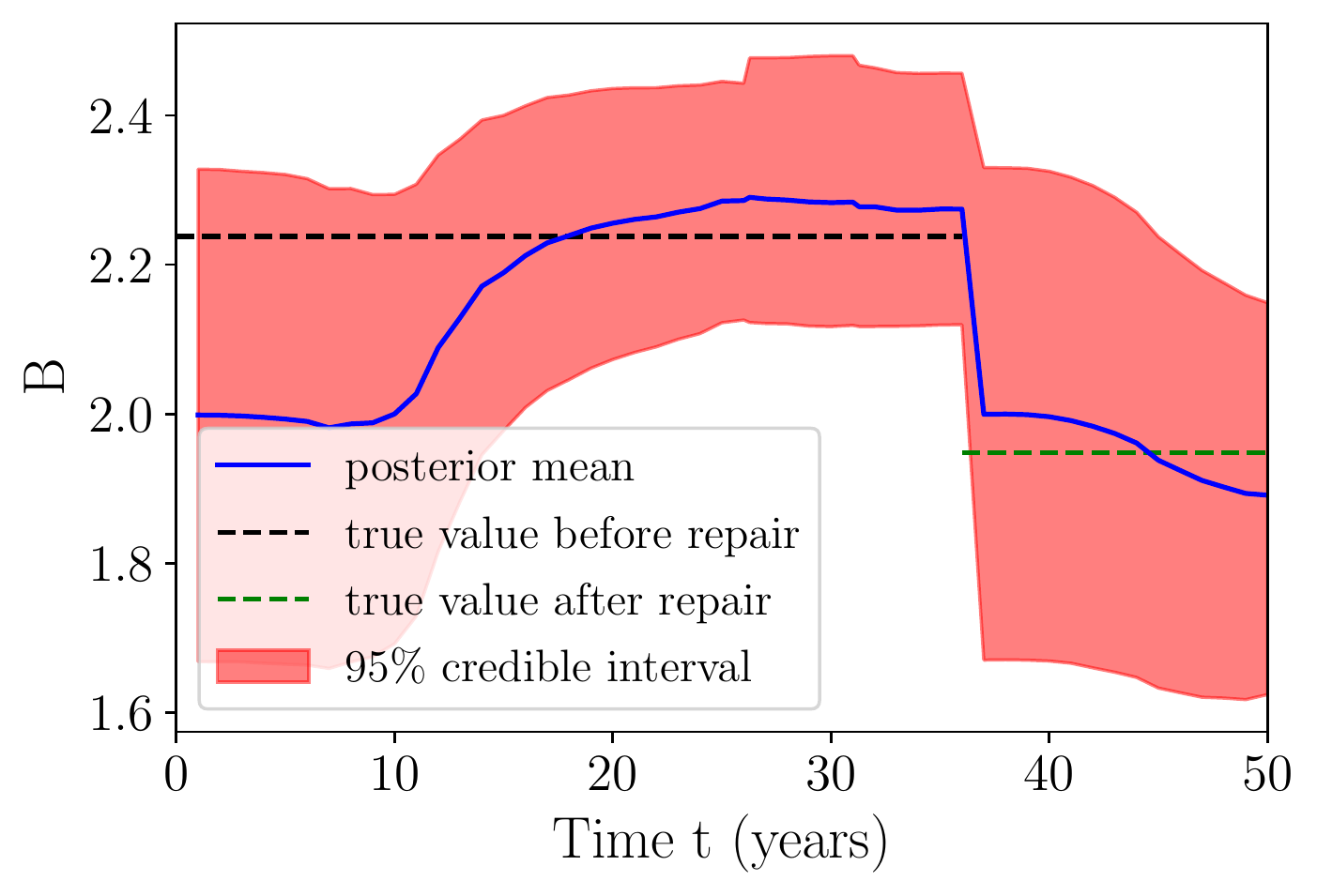}  
		\end{subfigure}
		\caption{Bayesian filtering of the deterioration state, the time-invariant model parameters and the reliability using continuous SHM and inspection data. In the bottom two panels, $A$ and $B$ correspond to the time-invariant parameters of the deterioration model \eqref{deterioration_model}}.
		\label{f:1st_case_study_SHM_2}
	\end{figure}

	Figures \ref{f:1st_case_study_inspect_2}, \ref{f:1st_case_study_SHM_2} plot the sequential Bayesian estimates for a second potential underlying ``true" deterioration process realization. Without SHM, inspections are performed at $\Delta t_i$=5 year intervals, when $p_{th}^I$ is exceeded, as well as when an extreme event occurs. More specifically, for the realization in Figure \ref{f:1st_case_study_inspect_2}, $t_{insp}=[5,10,15,20,25,26.3,31,31.2,36,42,47]$. A repair is performed at year 36, after an inspection triggered by $p_{th}^{I}$ at year 36 causes the $p_{th}^{R}$ to be exceeded. With SHM instead, $t_{insp}=[26.3,31.2,32,33, 34, 35,36]$ (this shows that the heuristic $\boldsymbol{w}_2^*$ is indeed suboptimal for the case with SHM). The inspections to complement the SHM are triggered when an extreme event occurs, and when $p_{th}^{I}$ is exceeded. Also with SHM, a repair at year 36 is informed by $p_{th}^{R}$. For this second realization, $C_{\text{tot}}\left(\boldsymbol{x}, \boldsymbol{z}_{insp}, \boldsymbol{w}_1^*\right) - C_{\text{tot}}\left(\boldsymbol{x}, \boldsymbol{z}_{insp}, \boldsymbol{z}_{SHM}, \boldsymbol{w}_2^*\right)=6.02\cdot10^4$\euro. 
	
	In Figure \ref{f:1st_case_study_SHM_2}, the bottom two panels further plot the estimates of the time-invariant gradual deterioration parameters $A, B$. Learning deterioration model parameters, and quantifying their uncertainty, is instrumental for predictive maintenance tasks.
	
	In this subsection, we illustrated the sequential decision making process and the corresponding life-cycle cost calculation for two samples of the underlying ``true" deterioration process and the corresponding sampled inspection and SHM data. Since the quantification of the $VoSHM$, as shown in Equation \eqref{VoSHM_heuristic}, requires the evaluation of the expectation operator, one needs to draw a sufficiently large finite number of samples of $\boldsymbol{X}$ and $\boldsymbol{Z}_{insp}/ \boldsymbol{Z}_{SHM}$ and compute the cost difference $C_{\text{tot}}\left(\boldsymbol{X}, \boldsymbol{Z}_{insp}, \boldsymbol{w}_1^*\right) - C_{\text{tot}}\left(\boldsymbol{X}, \boldsymbol{Z}_{insp}, \boldsymbol{Z}_{SHM}, \boldsymbol{w}_2^*\right)$ for each of those, and then take the mean value. For this first case study, with $n_{MCS}=1000$ samples, $VoSHM = 1.11\cdot10^5$\euro, which indicates a potential benefit of installing an SHM system on the deteriorating bridge structure. This value does not contain the total life-cycle cost of the SHM system itself (cost of installation, maintenance, repair, etc.). One should compare the $VoSHM$ value with the expected total life-cycle cost of the SHM system, and then decide whether installing such a system can be cost-beneficial. Furthermore, the presented VoSHM analysis is based on the premise that the SHM system will be continuously operating in an unobstructed fashion, and that complete trust will be put on the SHM system to replace all periodic inspections, which seems unrealistic. In this regard, the obtained $VoSHM$ value provides an upper limit to the potential economic benefit that uninterrupted monitoring with the investigated SHM system would generate.
	
	\subsubsection{Second case study: Gradual deterioration and shock deterioration due to an unobserved event}
	\label{subsubsec:Second}
	
	As a second case study, we assume scenarios when the CPP shock deterioration occurs due to an unobserved event, e.g., due to a sudden bridge bearing failure, or due to an extreme truck overloading. In such cases, without continuous SHM, no visual inspection will take place at the time of occurrence of this extreme event, since there is no knowledge of this occurrence. In contrast, with continuous vibration-based SHM in place, one will have access to continuous vibration data, and therefore data will be available right after the deterioration jump occurrence. This can lead to a timely tracking of the deterioration state increment and a subsequent prompt inspection or repair decision.
	
\begin{figure}[htp]
		\centering
		\begin{subfigure}{.40\textwidth}
			\centering
			\includegraphics[width=1.\linewidth]{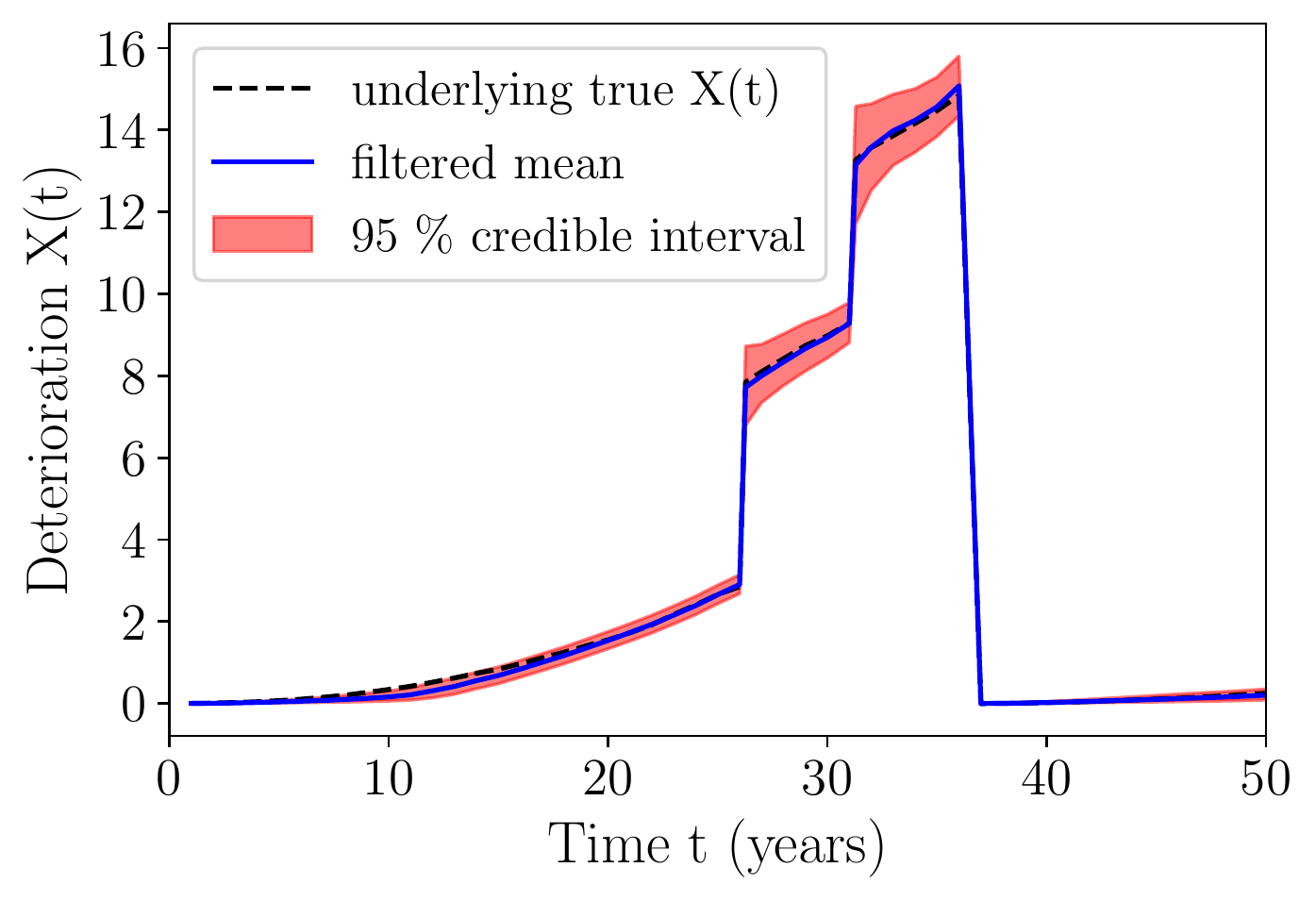}  
		\end{subfigure}
		\begin{subfigure}{.415\textwidth}
			\centering
			\includegraphics[width=1.\linewidth]{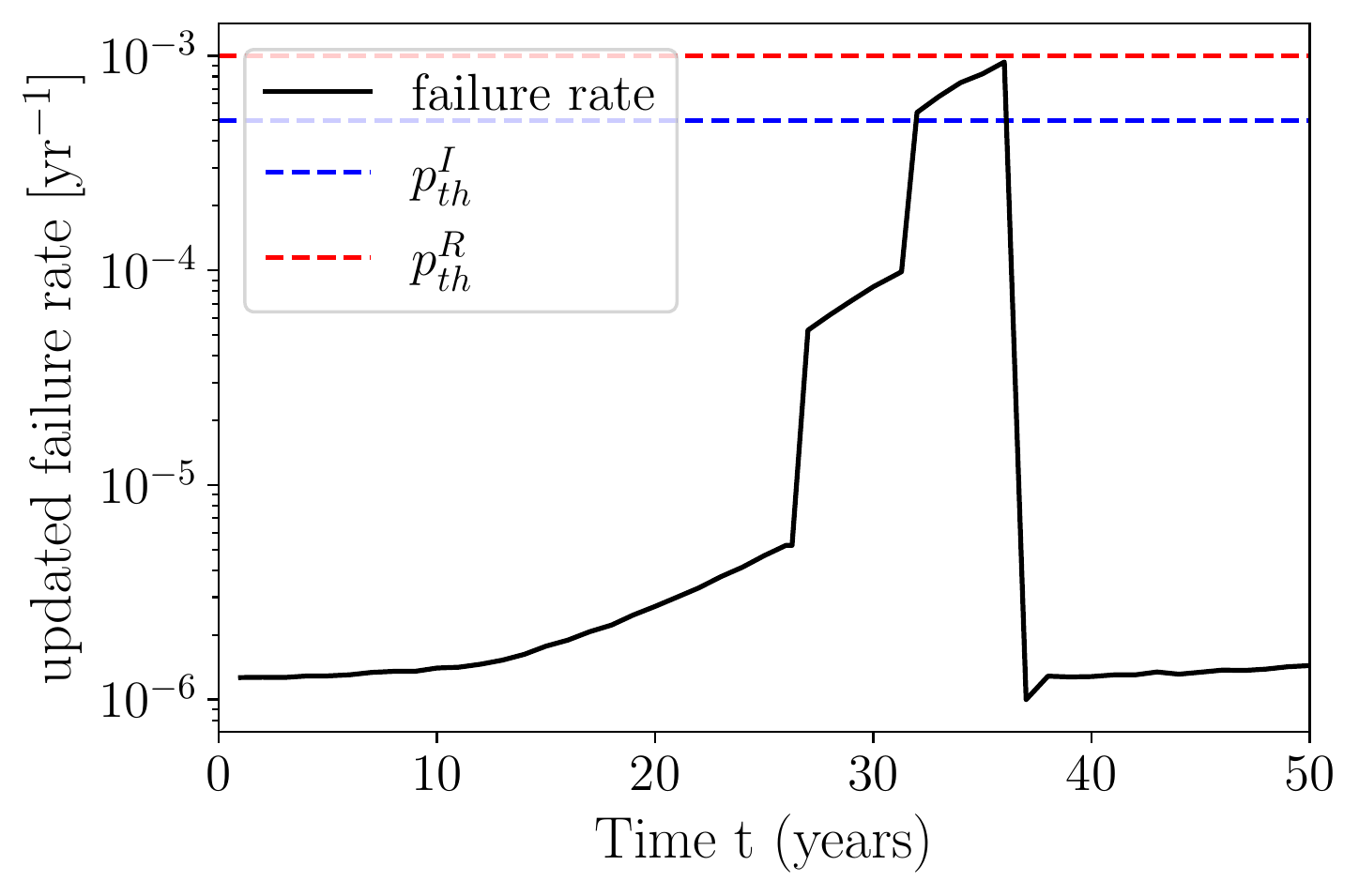}
		\end{subfigure}
		\caption{Bayesian filtering of the deterioration state and the reliability using continuous SHM and inspection data.}
		\label{f:2nd_case_study_SHM}
	\end{figure}

\begin{figure}[htp]
		\centering
		\begin{subfigure}{.40\textwidth}
			\centering
			\includegraphics[width=1.\linewidth]{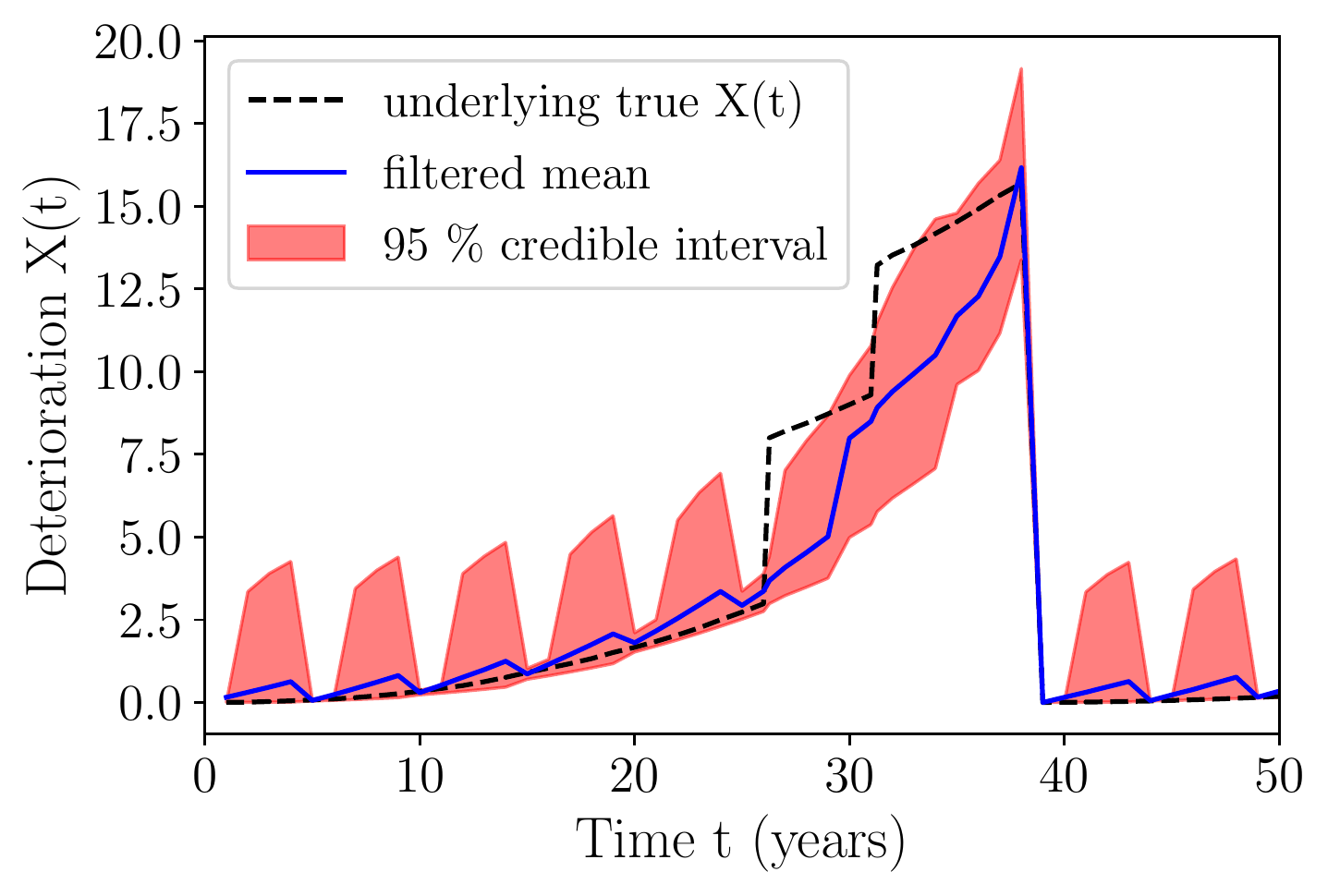}
		\end{subfigure}
		\begin{subfigure}{.415\textwidth}
			\centering
			\includegraphics[width=1.\linewidth]{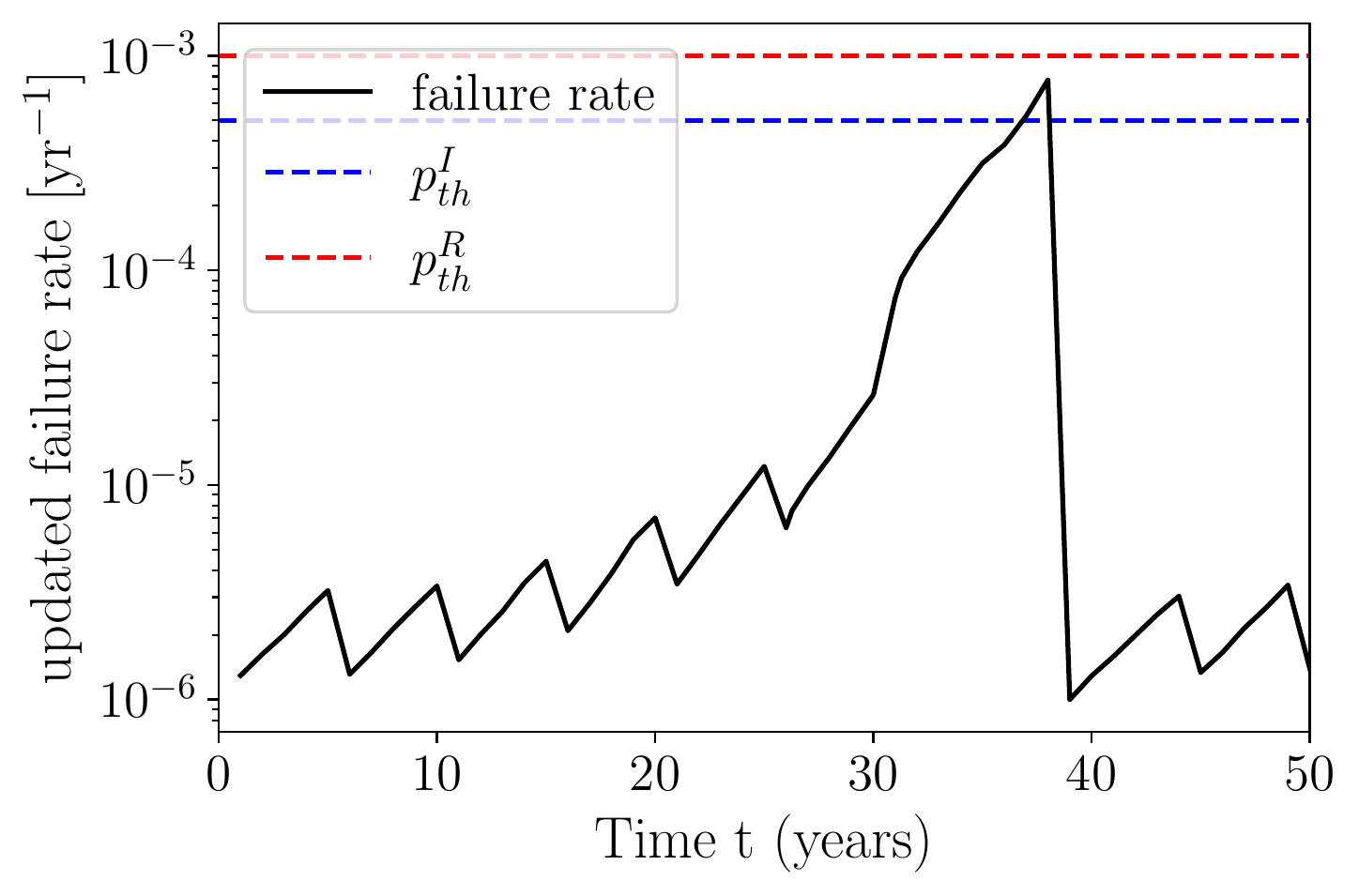}  
		\end{subfigure}
		\caption{Bayesian filtering of the deterioration state and the reliability using intermittent inspection data.}
		\label{f:2nd_case_study_inspect}
	\end{figure}

	We consider the same single underlying true realization of the deterioration process as in Figures \ref{f:1st_case_study_inspect_2}, \ref{f:1st_case_study_SHM_2}, and demonstrate the corresponding sequential decision making in the second case study. Comparing the left panel of Figure \ref{f:2nd_case_study_SHM} to the top left panel of Figure \ref{f:1st_case_study_SHM_2}, one observes that the estimation with SHM barely differs between the two case studies, due to the continuity of the SHM data. The actions corresponding to Figure \ref{f:2nd_case_study_SHM} are inspections at $t_{insp}=[32,33, 34, 35,36]$, as informed by exceedance of $p_{th}^I$, and repair at $t_{rep}=36$. On the other hand, comparing Figure \ref{f:2nd_case_study_inspect} to Figure \ref{f:1st_case_study_inspect_2}, it is clear that the absence of visual inspection measurements right after the deterioration jump events leads to a severe underestimation of the underlying true deterioration state. For the same heuristic thresholds as the ones in Figure \ref{f:1st_case_study_inspect_2}, the actions corresponding to Figure \ref{f:2nd_case_study_inspect} (without SHM) are inspections at $t_{insp}=[5,10,15,20,25, 30, 35, 37, 38,44,49]$ and repair at $t_{rep}=38$, i.e., two years after the repair informed by SHM, with an increased risk of failure during this time period. For this single realization, in the second case study, $C_{\text{tot}}\left(\boldsymbol{x}, \boldsymbol{z}_{insp}, \boldsymbol{w}_1^*\right) - C_{\text{tot}}\left(\boldsymbol{x}, \boldsymbol{z}_{insp}, \boldsymbol{z}_{SHM}, \boldsymbol{w}_2^*\right)=1.14\cdot10^5$\euro. 
	
	With $n_{MCS}=1000$ samples, in the second case study we quantify $VoSHM = 1.42\cdot10^5$\euro, which is larger than the $VoSHM$ result obtained for the first case study. This increased $VoSHM$ value is expected.
	
	\subsubsection{Third case study: Near-real time VoSHM in case of extreme event}
	\label{subsubsec:Third}

	As discussed in Section \ref{subsec:Near real-time}, SHM can be valuable in informing near real-time decisions for avoiding catastrophic failures, or avoiding unnecessary close-downs after the occurrence of an extreme event. As a third case study, we consider such an SHM for near-real time decision support.
	
	As in the first case study of section \ref{subsubsec:First}, we assume that a CPP shock deterioration corresponds to an observed extreme event. When such an event occurs, e.g., a flood, an inspection will have to take place right after the event, both in the case with and without SHM. However, we consider that it will only be possible to perform the inspection one week after the event, e.g., because inspectors cannot operate until the water level reduces or because they do not have the capacity to inspect all bridges in a short time. In this scenario, without SHM, an operator will have to close down the bridge for this one week until the inspection. This decision will induce a corresponding cost. According to \cite{Zhu_2010}, a close-down of a road bridge in the USA can cost up to $2\cdot10^5$ \$/day, while according to \cite{Lamb_2019, Sasidharan_2021}, in the UK scour-related close-downs of a railway bridge can induce costs up to $1.65\cdot10^5\text{\pounds}$/day. Herein, we assume a cost of close-down $c_{clsdn}=1.5\cdot10^5$\euro/day. This cost is subject to discounting.
	
	The benefit of continuous SHM is that the SHM system will be in operation also right after the extreme event, timely providing vibrational data from the state of the structural system. Already with the near-real time SHM data (hours) after the extreme event occurrence, the bridge operator will have an updated estimate of the state, and can therefore use this estimate to decide whether to close down the bridge for one week until the inspection, or whether to continue operating the bridge. We assume that in presence of the SHM data, the operator will be willing to accept a risk associated with continuation of the bridge operation if the updated estimate of the failure rate is lower than $p_{I}^{th}$. If $p_{I}^{th}$ is exceeded, then the bridge will be closed down, with an associated cost of $c_{clsdn}=1.5\cdot10^5\text{\euro}$/day.
	
	To summarize, in the case without SHM, inspections take place every 5 years, when $p_{I}^{th}$ is exceeded, and one week after an extreme event occurrence. The bridge is always closed down for this one week until the inspection. A repair is triggered when $p_{R}^{th}$ is exceeded. 
	
	In the case with SHM, inspections take place when $p_{I}^{th}$ is exceeded,  and one week after an extreme event occurrence. If the near real-time SHM-informed estimate of the failure rate exceeds $p_{I}^{th}$, then the bridge is closed down for this one week until the inspection, otherwise it is allowed that the bridge continues operating. A repair is triggered when $p_{R}^{th}$ is exceeded.
	
	With the setup described above, we run the VoSHM analysis with $n_{MCS}=1000$ samples, and we obtain an estimate of the  near-real time $VoSHM=1.34\cdot10^{6}$\euro. The $VoSHM$ in this third case study is one order of magnitude larger than in the previous two case studies. The reason for this is that the close-down cost per day is large, and hence the expected total close-down costs largely dominate the expected total life-cycle cost (the are one order of magnitude larger than the expected total inspection, repair, and risk costs). 
	
	This result shows that an effective SHM system for near real-time diagnostics might provide significant economic benefit, if it leads to avoidance of unnecessary close-downs.
	
	\subsubsection{Fourth case study: Reliability-based management}
	
	In all three case studies \ref{subsubsec:First}-\ref{subsubsec:Third}, we employ the heuristic parameter vector $\boldsymbol{w}_1^*=[p_{th}^{I*} = 5\cdot10^{-4}, p_{th}^{R*}=1\cdot10^{-3}, \Delta t_I=5]$, which is optimized for the case without SHM, and $\boldsymbol{w}_2^*=[p_{th}^{I*}= 5\cdot10^{-4}, p_{th}^{R*}=1\cdot10^{-3}, \Delta t_I=\infty]$. The values $p_{th}^{I*}$ and $p_{th}^{R*}$ of the thresholds used for inspection and repair decisions imply failure rates that are higher than what is typically accepted by authorities. As mentioned in Section \ref{subsec:heuristics}, the presented framework can also be used when replacing the optimization of the heuristic parameters by a choice based on expert assessment. The current subsection aims to demonstrate the VoSHM quantification in the case when reliability thresholds are imposed by authorities.
	
	To this end, we revisit the first case study of Section \ref{subsubsec:First}, but we now assume that a threshold for repairs $p_{th}^{R}=10^{-5}$ is imposed, and not subject to any optimization. We therefore optimize only the inspection threshold, and we find $p_{th}^{I*}=7\cdot10^{-6}$. We eventually use the heuristic parameter vector $\boldsymbol{w}_1^*=[p_{th}^{I*} = 7\cdot10^{-6}, p_{th}^{R}=10^{-5}, \Delta t_I=5]$ for the case of inspections only, and $\boldsymbol{w}_2^*=[p_{th}^{I*} = 7\cdot10^{-6}, p_{th}^{R}=10^{-5}, \Delta t_I=\infty]$ for the case with SHM, and we run a VoSHM analysis with $n_{MCS}=1000$ samples. We quantify $VoSHM=7.70\cdot10^{4}$\euro. This value is slightly lower than the VoSHM value from the first case study of Section \ref{subsubsec:First}, where both thresholds $p_{th}^{I}, p_{th}^{R}$ were optimized for the case without SHM.

	\section{Concluding remarks}
	This paper presents a Bayesian decision analysis framework for quantifying the expected gains that continuous vibration-based SHM-aided maintenance planning can provide when compared against the currently dominant approach of intermittent inspection-based maintenance planning; the Value of SHM (VoSHM) metric is adopted for formally computing this benefit. The framework requires the a-priori definition of damage scenarios and the associated stochastic deterioration models, describing the damage evolution over a target structure's lifetime. Furthermore, a case-specific SHM system model is necessary to allow for sampling of monitoring information. Contingent on these considerations, this framework can support the a-priori decision, in an operational evaluation level, on whether opting for an SHM system on a target structure can provide economic benefit.
	
	We detail the computational aspects of a VoSHM analysis, which involves stochastic sequential decision making, Bayesian analysis and structural reliability analysis. The modal data, identified sequentially over the structural life-cycle at different damage levels and for varying environmental conditions, is sampled in a realistic manner, following a state-of-the-art operational modal analysis procedure. The effect of the environmental variability present in the identified modal data is accounted for via Bayesian analysis. Sequential Bayesian updating of the deterioration state and model parameters, and consequently the structural reliability, is efficiently performed via adoption of a particle filtering scheme. Heuristic decision strategies, based on the updating of the risk estimation through inspection and monitoring, simplify the computationally challenging solution to the stochastic sequential decision making problem. A detailed algorithmic summary of a VoSHM analysis is provided, which is meant to act as an implementation template of the proposed framework for the interested reader.
	
	We discuss a novel classification of SHM use cases in terms of the associated time scales for decision making for infrastructure management. A gradual and shock stochastic deterioration model is employed, which is flexible in simulating various of these use cases. By means of a numerical model of a deteriorating two-span bridge system, we showcase the VoSHM analysis for four different case studies, across different time scales. The results show that investing in SHM systems can potentially lead to large benefits. It should be noted that the VoSHM framework of this paper does not incorporate deterioration or failure of the SHM system itself, and does not take into account modeling errors; it thus provides an upper limit to the ``true" VoSHM. For the purpose of illustration, in the presented numerical investigations, only a single deteriorating component has been considered, which tends to underestimate the VoSHM. The framework can, however, be extended to multiple deteriorating components. In general, the models utilized to showcase the illustrated framework in this paper can be extended to incorporate additional factors and uncertainties. While this can lead to increased computation, the main challenge to such extensions lies in the need for more detailed modeling, which is often difficult to justify in real-world projects. Hence, we believe that the level of detailing and accuracy reflected in the illustration in this paper is representative of what one can do when assessing the benefit of real-world SHM systems.
	
	This framework can be applied for optimal sensor placement studies as well. In this regard, the sensor placement which leads to the optimal balance between a large VoSHM and a low life-cycle cost of the SHM system would be the preferred arrangement.

	\section*{Declaration of competing interest}
	The authors declare that they have no known competing financial interests or personal relationships that could have appeared to influence the work reported in this paper.
	
	\section*{Acknowledgments}
    The work of A. Kamariotis and E. Chatzi has been carried out with the support of the Technical University of Munich - Institute for Advanced Study, Germany, funded by the German Excellence Initiative and the T{\"U}V S{\"U}D Foundation. We thank Dr. Konstantinos Tatsis for providing the code for the stochastic subspace identification algorithm and for the support provided for seamless integration of the SSI algortihm within the presented framework.
	
	\bibliography{mybibfile}
	
\end{document}